\documentclass{aa}  

\usepackage{graphicx}
\usepackage[varg]{txfonts}

\newcommand{\nperlc}{165}
\newcommand{\nper}{176}
\newcommand{\nperact}{113}
\newcommand{\npertotal}{279}

\begin{document} 

   \title{Rotation periods for cool stars in the open cluster NGC\,3532
		  \thanks{Based on observations at Cerro Tololo Inter-American Observatory, National Optical Astronomy Observatory under proposal 2008A-0476.}
	  \thanks{The full Table 2 is only available in electronic form
	  	at the CDS via anonymous ftp to cdsarc.u-strasbg.fr (130.79.128.5)
	  	or via http://cdsweb.u-strasbg.fr/cgi-bin/qcat?J/A+A/}}

   \subtitle{The transition from fast to slow rotation}

   \author{D. J. Fritzewski\inst{1}
          \and
          S. A. Barnes\inst{1, 2}
          \and
          D. J. James\inst{3, 4}
          \and
          K. G. Strassmeier\inst{1}
          }

   \institute{Leibniz-Institut f\"ur Astrophysik Potsdam (AIP),
              An der Sternwarte 16, 14482 Potsdam, Germany\\
              \email{dfritzewski@aip.de}
          \and
              Space Science Institute, 4750 Walnut St., Boulder, CO 80301, USA
              \and
              Center for Astrophysics $\vert$ Harvard \& Smithsonian, 60 Garden Street, Cambridge, MA 02138, USA
		  \and
		      Black Hole Initiative at Harvard University, 20 Garden Street, Cambridge, MA 02138, USA
		      }

   \date{}

 
  \abstract
   {A very rich cluster intermediate in age between the Pleiades (150\,Myr) and the Hyades (600\,Myr) is needed to probe the rotational evolution, especially the transition between fast and slow rotation that occurs between the two ages.
   }
   {We study the rich 300\,Myr-old open cluster NGC\,3532 to probe this important transition and to provide constraints on angular momentum loss. 
   Measuring the rotation periods builds on our prior work of providing spectroscopic membership information for the cluster, and it supports the chromospheric activity measurements of cluster stars that we provide in a companion paper.
   }
   {Using 42\,d-long photometric time series observations obtained with the Yale 1\,m telescope at CTIO, we measured rotation periods for members of NGC\,3532 and compared them with the predictions of angular momentum evolution models.
   }
   {
   We directly measured \nper{} photometric rotation periods for the cluster members. 
   An additional \nperact{} photometric rotation periods were identified using activity information, described fully in the companion paper, 
   resulting in a total sample containing \npertotal{} rotation periods for FGKM stars in NGC\,3532. 
   The colour-period diagram constructed from this rich data set shows a well-populated and structured slow rotator sequence, and a fast rotator sequence evolved beyond zero-age main sequence age whose stars are in transition from fast to slow rotation.
   The slow rotator sequence itself is split into slightly slower and faster rotators, a feature we trace to photometric binary status.
   We also identify an extended slow rotator sequence extending to $P \sim 32$\,d, apparently the analogue of the one we previously identified in NGC\,2516.
   We compare our period distribution to rotational isochrones in colour-period space and find that all considered models have certain shortcomings.
   Using more detailed spin-down models, we evolve the rotation periods of the younger NGC\,2516 forward in time and find that the spindown of the models is too aggressive with respect to the slow rotators. 
   In contrast, stars on the evolved fast rotator sequence are not spun down strongly enough by these models. 
   Our observations suggest a shorter crossing time for the rotational gap, one we estimate to be $\sim 80$\,Myr for early-K\,dwarfs.
   }
  {}


   \titlerunning{Rotation periods for cool stars in the open cluster NGC\,3532}
   \authorrunning{D. J. Fritzewski et al.}
   \maketitle
%
\section{Introduction}
The vast majority of solar-type stars in young open clusters rotate either fast ($P_\mathrm{rot}\lesssim2$\,d) or slowly ($P_\mathrm{rot}\gtrsim3$\,d), forming two distinct, mass-dependent groups.
These groups often appear as sequences when cluster rotation periods are plotted against stellar colour or an equivalent variable.
Over their early main-sequence lifetime (meaning between Pleiades and Hyades age) all FGK stars apparently converge onto the latter group, a narrow sequence of slow rotators, erasing the evidence of the prior rotational evolution \citep{2003ApJ...586..464B}. 
While the existence and evolution of the slow rotators themselves has been explored extensively both observationally (e.g. \citealt{1987ApJ...321..459R, 2015Natur.517..589M, 2016ApJ...823...16B, 2020A&A...644A..16G}) and theoretically (e.g. \citealt{2010ApJ...722..222B, 2015ApJ...799L..23M, 2020A&A...636A..76S}), measuring the transformation of stars from fast- to slow rotators and understanding the characteristics of that transition demands special effort.

Such work requires the study of the richest-available open clusters, excellent membership information, supporting spectroscopic studies, and of course similar sensitivity to rotation periods,
allowing both rotational branches and more generally the entire relevant rotational distributions to be populated.
We previously published a detailed study of the rich 150\,Myr-old zero-age main sequence (ZAMS) open cluster NGC\,2516 \citep{2020A&A...641A..51F}, based on a 42\,d observing baseline.
Here, we present an equivalent study of the rich 300\,Myr-old open cluster NGC\,3532 which
provides the best opportunity for sampling the ZAMS-to-Hyades age transition.
NGC\,3532 was observed in the same observing run, in parallel with NGC\,2516, essentially assuring comparable rotation period sensitivity.  

The first fast rotators ($P_\mathrm{rot} \sim 0.5$\,d) were discovered in the Pleiades \citep{1982Msngr..28...15V,1987A&AS...67..483V}. 
Shortly thereafter, \cite{1987ApJ...318..337S} showed that they cannot spin down with the same rate as the slow rotators, whose 
angular momentum loss rate ($\mathrm{d}J/\mathrm{d}t$)
scales with the cube of the angular velocity ($\Omega$): $\mathrm{d}J/\mathrm{d}t\propto -\Omega^{3}$ \citep{1988ApJ...333..236K}. 
Such a strong dependence on the angular velocity would brake the fast rotators in the Pleiades on timescales shorter than their main sequence age and one would not observe them at all. 
In fact, fast rotators can also be observed in older open clusters. 
To resolve this issue, the idea emerged of a spin-down that depends linearly (rather than cubically) on the angular velocity \citep{1991ApJ...376..204M, 1993ApJ...409..624S, 1995ApJ...441..865C}. 
In this framework, the large-scale magnetic field is thought to saturate and the spin-down becomes less efficient. 
\cite{2019ApJ...876..118S} observe this saturation for the fastest rotators, but even with this evidence it is not understood how the transition from fast to slow rotation occurs and whether it is stochastic \citep{2014ApJ...789..101B} or driven by a different magnetic field configuration \citep{2003ApJ...586..464B,2018ApJ...862...90G,2019ApJ...886..120S}.

Rotation periods are the primary observable of the stellar angular momentum content. We measure rotation periods from starspot-modulated photometric light curves \citep{1947PASP...59..261K, 1987A&AS...67..483V, 2009A&ARv..17..251S}. Such photometric rotation periods are routinely obtained with high precision for low-mass pre- and main sequence stars of all ages (e.g. \citealt{2005ApJ...633..967H,2016ApJ...823...16B}). 
Stars in open clusters are of particular interest because such populations can be arranged in an age-ranked sequence, providing valuable information in understanding the stellar spin-down.

At the zero-age main sequence five open clusters
[NGC\,2516 \citep{2007MNRAS.377..741I, 2020A&A...641A..51F}, 
M\,35 \citep{2009ApJ...695..679M},
M\,50 \citep{2009MNRAS.392.1456I},
Pleiades \citep{2010MNRAS.408..475H, 2016AJ....152..113R},
and Blanco\,1 \citep{2014ApJ...782...29C,2020MNRAS.492.1008G}] 
with measured rotation periods provide a cornerstone for the angular momentum evolution of cool main sequence stars. 
In \cite{2020A&A...641A..51F} (hereafter F20), we have shown that these clusters host a universal rotation period distribution which confirms their isochrone-based ages, and facilitates the reliable age ranking of other open clusters. 
M\,34 at 220\,Myr \citep{2006MNRAS.370..954I,2010A&A...515A.100J,2011ApJ...733..115M} provides the first evolutionary step beyond the zero-age main sequence.
However, M\,34 is not a particularly rich open cluster, and its age of 220\,Myr is not advanced enough to observe significant rotational evolution unambiguously.

Significant further rotational evolution can be observed in the 300\,Myr-old NGC\,3532 \citep{N3532RV}. 
Not only is it unique in age among the nearby open clusters, but it provides a very large stellar population. 
\cite{N3532RV} showed that it is one of the richest open clusters within the 500\,pc horizon, enabling the identification and study of details of stellar rotation not visible elsewhere, and in particular the evolution of the fast rotators at this key age. 
As part of our efforts on this rich southern open cluster, potentially a benchmark object, we have already provided 
membership information (\citealt{N3532RV}, hereafter F19) based on radial velocity observations, 
multi-colour photometry by \cite{clem},
and \emph{Gaia} astrometry. 
This membership work provides a strong foundation for a careful study of the rotation rates (really, periods) of its cool star members in the present paper. 
Furthermore, we analyse the chromospheric activity and its connection to stellar rotation in a companion study (\citealt{N3532act}, hereafter F21act).
In consequence, this uniquely-aged, rich open cluster provides a snapshot of the on-going evolution from fast to slow rotation and, together with the period distributions in the group of younger open clusters, provides empirical constraints on the time-scale of the transition from fast to slow rotation.

NGC\,3532 is located in a crowded Galactic field in Carina at a distance of ${\sim}484$\,pc and has 
slightly sub-solar metallicity ([Fe/H] $=-0.07\pm0.1$, F19, [Fe/H]$_\mathrm{NLTE}=-0.10\pm0.02$, \citealt{2019A&A...628A..54K}). 
The interested reader can find further details and a historical overview in the introduction of F19.

The nearest older open cluster with measured rotation periods is M\,48 at 450\,Myr \citep{2015A&A...583A..73B}, followed by M\,37 (550\,Myr) \citep{2008ApJ...675.1254H}. 
The former is relatively sparse, and the latter is distant.
Very few FGK-type fast rotators are observed at such ages. 
By the age of the benchmark open cluster Hyades \citep{1987ApJ...321..459R, 2011MNRAS.413.2218D, 2016ApJ...822...47D, 2019ApJ...879..100D} and the similarly aged Praesepe \citep{2011MNRAS.413.2218D, 2017ApJ...839...92R} (both $\sim$600\,Myr), all solar-like stars have settled onto the slow rotator sequence.

Based on observational studies such as the ones summarised above, (semi-)empirical relations between the spin-down of the slow rotators and stellar age have been derived. 
\cite{1972ApJ...171..565S} was the first to show that $v_\mathrm{eq}\propto t ^ {-0.5}$ for solar-type stars 
(where $v_\mathrm{eq}$ is the equatorial rotation velocity and $t$ the stellar age).
This laid the foundation for quantitative relations between the rotation period and the stellar age, subsequently including an additional mass dependence. 
This gyrochronology, introduced by \cite{2003ApJ...586..464B} 
and re-calibrated later (Barnes 2007, \citealt{2008ApJ...687.1264M, 2009ApJ...695..679M, 2010ApJ...722..222B}), is a successful method to determine ages of slowly rotating stars \citep{2016ApJ...823...16B, 2020MNRAS.495L..61L}.
The ages of stars being otherwise difficult to obtain, the spin-down must be studied carefully if gyrochronology is to yield useful ages.

Fast rotators either might not be useful in determining stellar ages due to ambiguities, 
or else they may provide an additional sensitive probe to distinguish small age differences between otherwise similar open clusters.
Understanding their transition to slow rotation could also provide insights into the effect of planetary companions on angular momentum loss \citep{2010ApJ...723L..64C} 
and into the atmospheric evolution of those companions \citep{2015ApJ...815L..12J}.
Finally, 
the fast rotators are known to have activity behaviours that differ significantly from those of the slow rotators which tend to follow what is generally known as the rotation-activity relationship.
The measurement of their periods is essential to understanding their activity properties, as detailed in the companion paper (F21act).

This paper is structured as follows. 
In Sect.~\ref{sec:data}, we describe our observations, data reduction and photometry in relation to the previously obtained membership. 
The time-series analysis is carried out in Sect.~\ref{sec:timeseries}, yielding a large set of rotation periods. 
In the companion paper on stellar activity in NGC\,3532, we derive further activity-informed rotation periods from the same photometric time series.
We present and analyse the complete dataset in Sect.~\ref{sec:CPD}. 
In Sect.~\ref{sec:modelgyrochrones}, we compare the rotation period distribution in colour-period space to available rotational isochrones. 
Because of the 
inadequacies of the models in describing our NGC\,3532 data,
we compare the rotational distribution in Sect.~\ref{sec:detailedComp} directly to open clusters of bracketing ages, 
and also spin down our prior NGC\,2516 
rotational distribution to attempt to recreate the NGC\,3532 rotational distribution.
We also examine the transition from fast to slow rotation and estimate the transition timescale as well as the spin-down rate.
Finally, in Sect.~\ref{sec:conclusion}, we provide our conclusions.

\section{Time series photometry}
\label{sec:data}

\subsection{Observations}
\label{sec:observations}

We observed the Galactic open cluster NGC\,3532 from the Cerro Tololo Inter-American Observatory (CTIO) with the Yale 1\,m telescope operated by the SMARTS consortium between 19~February~2008 and 1~April~2008. 
Over the course of these 42\,d, we obtained time-series photometry without any loss due to weather. 
However, the observations contain a scheduling gap of four nights (from 8 March 2008 to 11 March 2008), slightly offset from the middle of our observing campaign.
The image scale of the \emph{Y4KCam} camera mounted at the CTIO Yale 1\,m telescope is $0\farcs289\,\mathrm{px}^{-1}$ which gives a $19.3\arcmin\times19.3\arcmin$ field of view with the $4064\,\mathrm{px}\times4064\,\mathrm{px}$ STA CCD detector. The average detector read noise was $4.8\,e^-$\,px$^{-1}$.

\begin{figure}
	\includegraphics[width=\columnwidth]{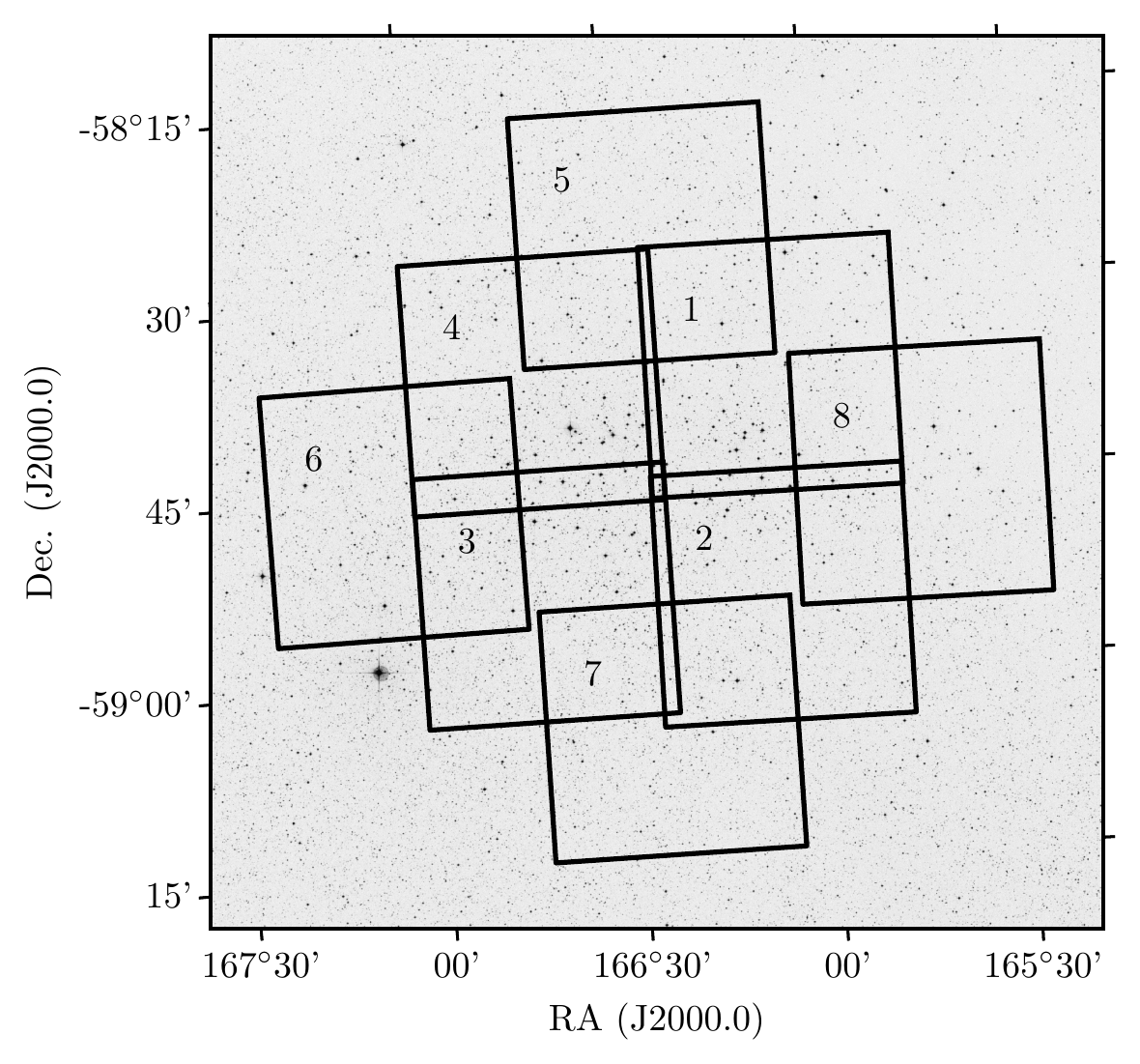}
	\caption{Field of NGC 3532, as seen in the Digitized Sky Survey 2 (red filter), with the individual observed fields indicated. The brightest cluster stars are located in the inner four fields (1 to 4). Their shorter exposure times were optimised for the solar-like stars. The outer fields (5 to 8) were exposed longer, with the fainter stars optimised. (Field numbers are positioned in the north-east corners.)}
	\label{fig:fields}
\end{figure}

We observed four inner fields of NGC\,3532 in both $V$ and $I_c$ filters, with exposure times of 120\,s and 60\,s respectively. 
We also observed four deeper outer fields for 600\,s in the $I_c$ filter.
The positions on sky of the eight fields are shown in Fig.~\ref{fig:fields} (centred at $\alpha=\mbox{11:05:39}$, $\delta=\mbox{-58:45:07}$ (J2000.0)). 
For the inner fields, we obtained ${\sim}120$ frames per field and for the deeper outer, fields ${\sim}95$ frames per field (see Table~\ref{tab:obslog}).
We note that the cluster extends significantly beyond even this relatively wide observing area.

The data presented in this work were obtained in the course of the same observing campaign as the observations of NGC\,2516 which are presented in F20. 
Consequently, the time baselines of the two photometric time series observations are identical, and the observing cadence is very similar, ensuring similar period sensitivity and comparability.
Further information about the observations and the data analysis is available in that publication.

\begin{table}
	\caption{Number of visits for all observed fields.}
	\label{tab:obslog}
	\begin{tabular}{l  l  c c  }
		\hline
		\hline
		Field name & Filter & Exposure time & Number of visits\\
		& & (s) &\\
		\hline
		F1 & $I_c$ & 60 & 120\\
		& $V$ & 120 & 119\\
		F2 & $I_c$ & 60 & 120\\
		& $V$ & 120 & 121\\			
		F3 & $I_c$ & 60 & 121\\
		& $V$ & 120 & 120\\
		F4 & $I_c$ & 60 & 120\\
		& $V$ & 120 & 117\\
		F5 (Deep North) & $I_c$ & 600 & 97\\
		F6 (Deep East) & $I_c$ & 600 & 90\\
		F7 (Deep South) & $I_c$ & 600 & 95\\		
		F8 (Deep West) & $I_c$ & 600 & 94\\
		\hline	 
	\end{tabular}
\end{table}

\subsection{Data reduction and photometry}
\label{sec:photometry}

We obtained zero-second bias images and per-filter sky-flat fields each night to enable routine calibration. 
A median bias frame was subtracted from all calibration and science images using IRAF\footnote{IRAF is distributed by the National Optical Astronomy Observatories, which are operated by the Association of Universities for Research in Astronomy, Inc., under cooperative agreement with the National Science Foundation.}. 
Additionally, we corrected the science data in each filter for pixel-to-pixel sensitivity differences using a per-filter balance frame. 
Dark current correction was not applied to the images because the dark current of \emph{Y4Kcam} is sufficiently low ($21\,e^-\,\mathrm{px}^{-1}\,\mathrm{h}^{-1}$).

Because the cluster is located at low Galactic latitude ($b=1.36^\circ$), the field of NGC\,3532 contains many background stars from the Galactic disc. 
In order to extract the best possible light curves from this crowded field, we performed PSF photometry with \textsc{DaoPhot~II} \citep{1987PASP...99..191S, 1994PASP..106..250S, 2003PASP..115..413S}. 
We decided to use a shared approach with both data sets (NGC\,2516 and NGC\,3532) to obtain comparable results. 
Hence, we followed the same workflow for the PSF photometry as summarised here, described in detail in F20, and to which we direct interested readers.

After initial source identification on the reference frame for each field, we selected the PSF stars. 
We matched all frames to this reference frame in the main photometric run to use the same set of PSF stars in each frame. 
We created a PSF model from this set for each image and used it to extract the stellar flux in \textsc{DaoPhot~II}. 
Thereafter, we used the \textsc{DaoMaster} software to cross-match the frames down to 0.3\,px and to create the individual light curves.

In total, we extracted ${\sim}200\,000$ light curves for ${\sim}110\,000$ individual stars. 
Stars with multiple light curves have either been observed in both filters, or in different fields, or in a combination of both.
These are retained and analysed separately to provide redundancy, as in F20.

\begin{figure}
	\includegraphics[width=\columnwidth]{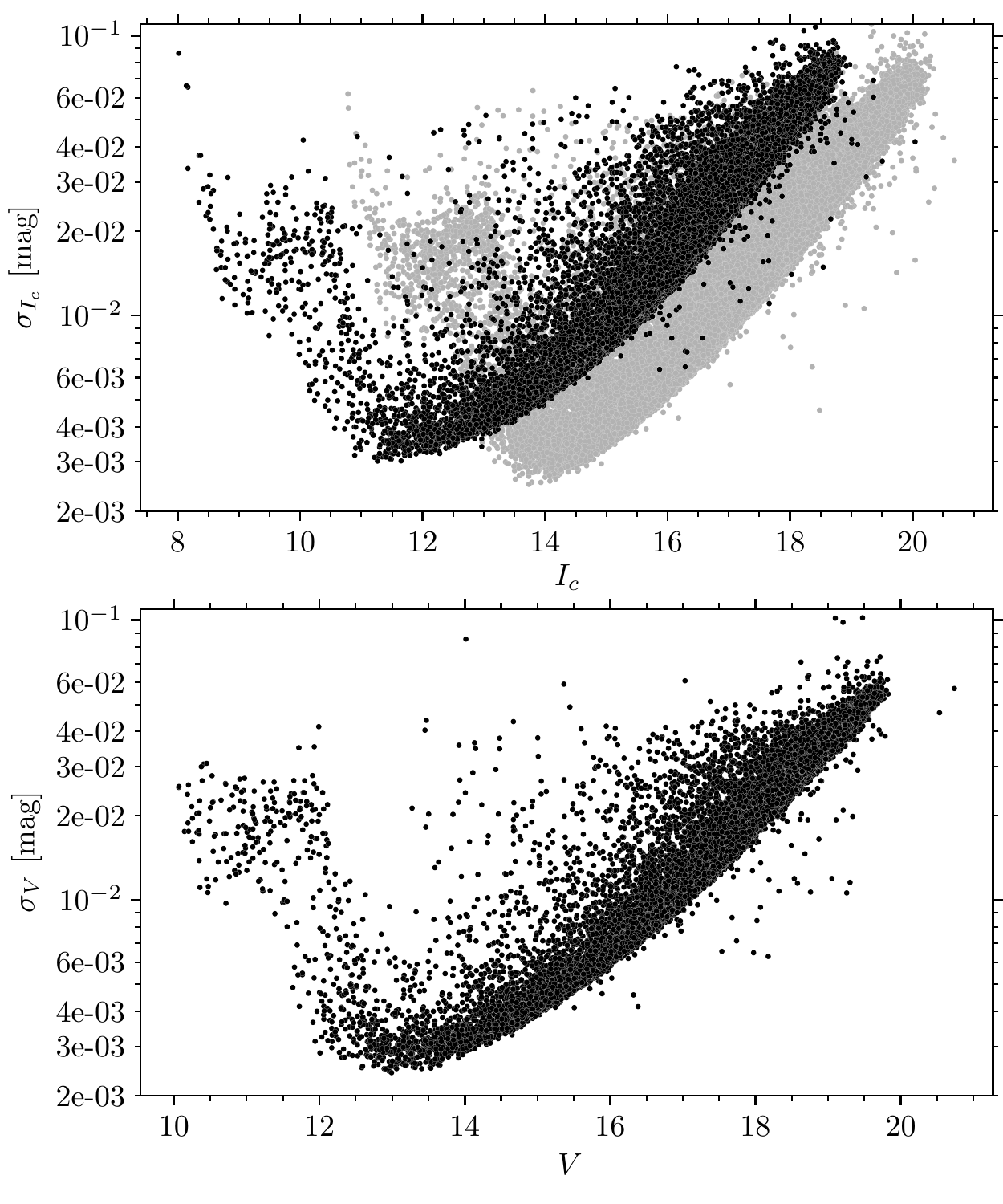}
	\caption{Photometric uncertainties ($\sigma$) in our time series observations for the relevant filter and exposure time combinations. \emph{Top:} Uncertainties for the two different $I_c$ exposure settings, 60\,s (black) and 600\,s (grey). \emph{Bottom:} Uncertainties for the 120\,s $V$ exposures. In combination, we are able to ensure a measurement precision better than 10\,mmag for a large fraction of the observed stars with time series.}
	\label{fig:photerror}
\end{figure}

The distribution of the mean photometric measurement uncertainties for the individual stars are shown in Fig.~\ref{fig:photerror} as a function of stellar magnitude for each of the filter and exposure time combinations\footnote{These values do not contain any contribution from stellar variability and should not be confused with the variability amplitude, whose analysis we defer to the companion paper (see Sect.~\ref{sec:varamp} for a brief summary.)}. 
The deep $I_c$ images (600\,s) have the smallest photometric uncertainties among all three configurations for stars fainter than $I_c=14$. 
In the range between $V=12.5$ and $V=14$, we achieved a $V$ band photometric accuracy better than 3\,mmag for a small fraction of the stars. 
For fainter stars, the photon noise-limited uncertainties are only slightly larger than for the deep $I_c$ photometry. 
Although both the deep $I_c$ and the $V$ observations have similar uncertainty distributions in their respective filters, the deep $I_c$ observations provide more precise photometry for a given star in the photon noise-limited case. 
For example, a cluster member at $I_c=17$ has an intrinsic colour of $(V-I_c)_0=2.5$. Hence, the deep $I_c$ observations are comparable to a $V=19.5$ star and we reach an accuracy $\sigma_{I_c}\leq0.01$ at this magnitude. 
The short (60\,s) $I_c$ photometry has larger uncertainties at the same $I_c$ magnitude, but is very valuable for the brighter stars. 
In F20, we analysed the photometric uncertainties in detail, providing confidence that the estimated values reflect the true measurement errors.

\subsection{Membership}

\begin{figure}
	\includegraphics[width=\columnwidth]{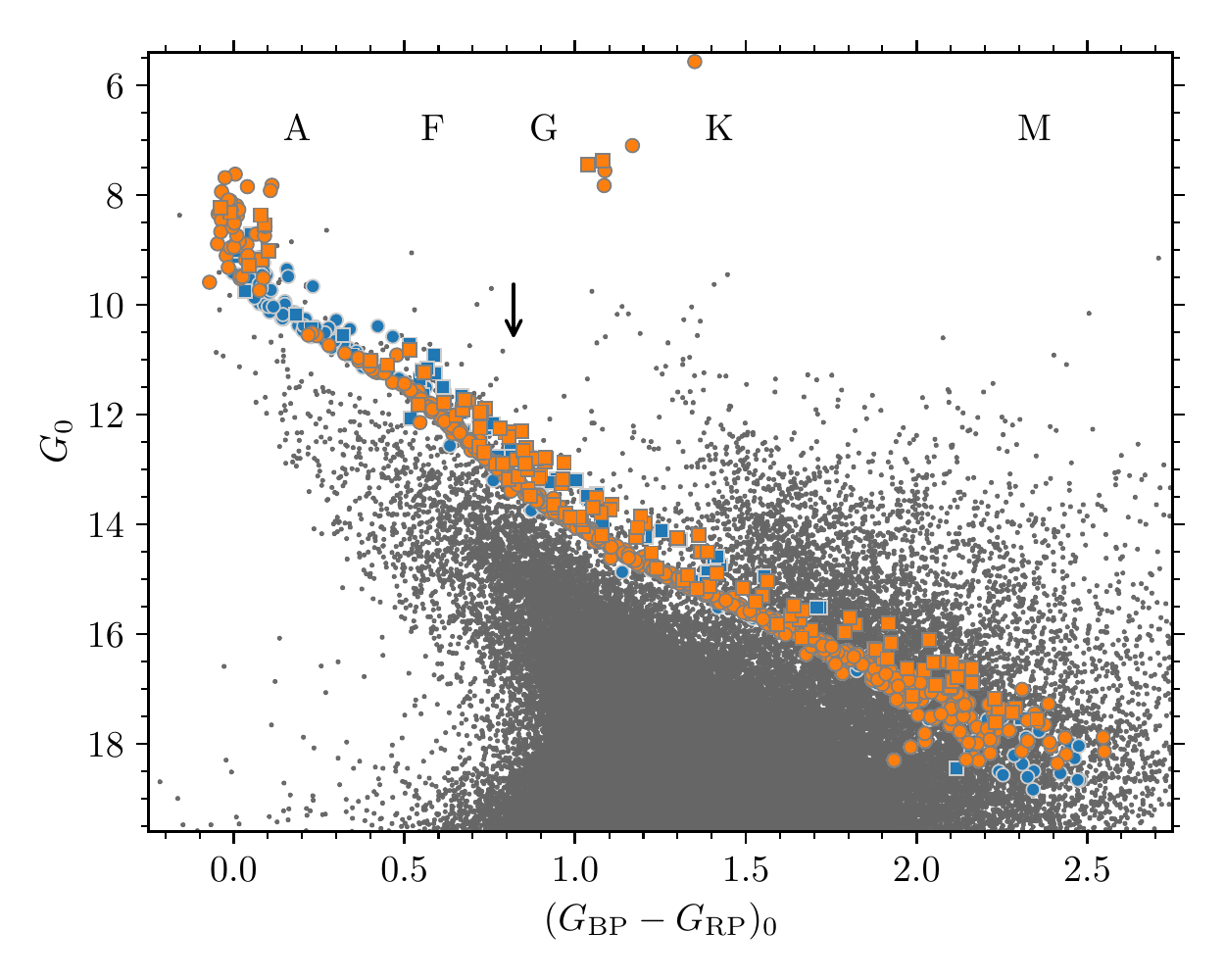}
	\caption{Colour-magnitude diagram in \emph{Gaia}~DR2 colour for NGC\,3532 with cluster members highlighted. Radial velocity members from \cite{N3532RV} are marked in orange and additional proper motion members in blue. The enormous field star population is shown in grey in the background. All data are dereddened with the average reddening value towards NGC\,3532 ($E_{G_\mathrm{BP}-G_\mathrm{BP}}=0.046$). Likely binaries are marked with squares. At the top, we indicate the different spectral types from \cite{2013ApJS..208....9P} for orientation. The arrow indicates the Solar colour \citep{2018MNRAS.479L.102C}. For colour-magnitude diagrams in other colours see Fig.~\ref{fig:CMD} below and Fig.~14 of F19.}
	\label{fig:memCMD}
\end{figure}

Well-defined open cluster members are crucial in constructing the clean, coeval, and chemically homogenous stellar samples needed to explore angular momentum evolution reliably. 
As part of our analysis of NGC\,3532, we have already provided such a membership determination in \cite{N3532RV}. 
There, we obtained radial velocity observations of ${>}1\,000$ photometric members of NGC\,3532 and found 660 exclusive member stars (with the inclusion of literature radial velocity data and \emph{Gaia}~DR2 astrometry).
Fig.~\ref{fig:memCMD} shows the cluster sequence in a colour-magnitude diagram (CMD) with Gaia~DR2 photometry \citep{2018A&A...616A...4E}. 
(CMDs in other filters, including $BV(RI)_c$, are presented in F19.) 
Although we were unable to survey the entire open cluster on account of its extent on sky, this set of bona-fide cluster members is much larger than for most other nearby open clusters. 
Hence, it is particularly well suited to the study of stellar rotation periods in NGC\,3532. 
In this work, we exclusively use these members, and conduct no additional membership analysis.

\subsection{Multiplicity}
\label{sec:binaries}

We also identify likely binaries among the cluster stars.
Unfortunately, the radial velocity binaries from F19, identified as part of our membership work, concern either stars brighter than the cool ones of interest here, or non-members.
However, we have identified likely binaries in NGC\,3532 using two methods.

Firstly, we classify all cluster members 0.25\,mag brighter than the stellar main sequence locus as potential binaries. 
In F19, we have provided this MS locus in various colours. 
The difference between the locus and the observed $G_0$ magnitude is the offset above the main sequence. 
We limit our search for stars in the range 
$0.5 < G_\mathrm{BP}-G_\mathrm{BP} < 2.3$ 
because (a) bluer stars are not of interest in our study of cool star rotation, and (b) for redder stars the photometric uncertainties become too large for meaningful results.
Secondly, as in prior studies (e.g. \citealt{2020MNRAS.496.1922B}), we use the \emph{Gaia} EDR3 \citep{2021A&A...649A...2L} renormalised unit weight error (RUWE) as a binarity criterion. 
We classify all stars with $\mathrm{RUWE}>1.2$ as astrometric binaries.

In total we identify 151 likely binaries among the radial velocity and proper motion members, corresponding to a binary fraction of 22\,\%. 
This is in reasonable agreement with the modelled binary fraction of 26.7\,\% of \cite{2020ApJ...901...49L} for NGC\,3532, although that too is probably an underestimate. 
More binaries will undoubtedly be identified with looser criteria, and of course additional data.
The combined set of binaries identified according to the above criteria is marked in the CMD in Fig.~\ref{fig:memCMD} and in other relevant diagrams below.

\section{Rotation periods}
\label{sec:timeseries}

Because this paper is restricted to a discussion of the rotation periods only of members of NGC\,3532 and because of the presence of a
large number of field stars in our photometry\footnote{The field around NGC\,3532 contains ${\gtrsim}100\,000$ sources of which ${\sim}1\%$ are possible cluster members.},
we confine our time series analysis only to cluster members of NGC\,3532.
We applied the same algorithm and the same pipeline for these data as we did for NGC\,2516 (F20) to construct the light curves from the time-series photometry, to analyse the time series, and to obtain rotation periods. 
Because the data were obtained in parallel, the light curves for both clusters are so similar in data structure that we were able to apply many tools developed for the one cluster without adjustments to the other cluster. 
The workflow presented below is described in greater detail in F20. 
(See especially the decision tree in Fig.~8 in that paper.)

\subsection{Initial periods}

As in F20, we applied five algorithms to determine the stellar rotation periods ($P_\mathrm{rot}$):
Lomb-Scargle \citep{1976Ap&SS..39..447L, 1982ApJ...263..835S}, in the form of the generalized Lomb-Scargle periodogram \citep{2009A&A...496..577Z}, 
the \textsc{clean} algorithm \citep{1987AJ.....93..968R, 2001SoPh..203..381C}, 
phase dispersion minimization (PDM, \citealt{1978ApJ...224..953S}), 
string length \citep{1983MNRAS.203..917D}, and the 
Gregory-Loredo periodogram \citep{1992ApJ...398..146G, 1999ApJ...520..361G}

After applying each algorithm to all light curves (examples of which are displayed in the Appendix in Fig.~\ref{appf:LCs}), we selected a common period following the scheme presented in F20 (summarised in Fig.~8 there). 
We examined each set of light curves and periodograms manually to assure ourselves 
that periodic variability is indeed present in the data, and to determine the common period from the (usually) multiple estimates.

\begin{figure}
	\includegraphics[width=\columnwidth]{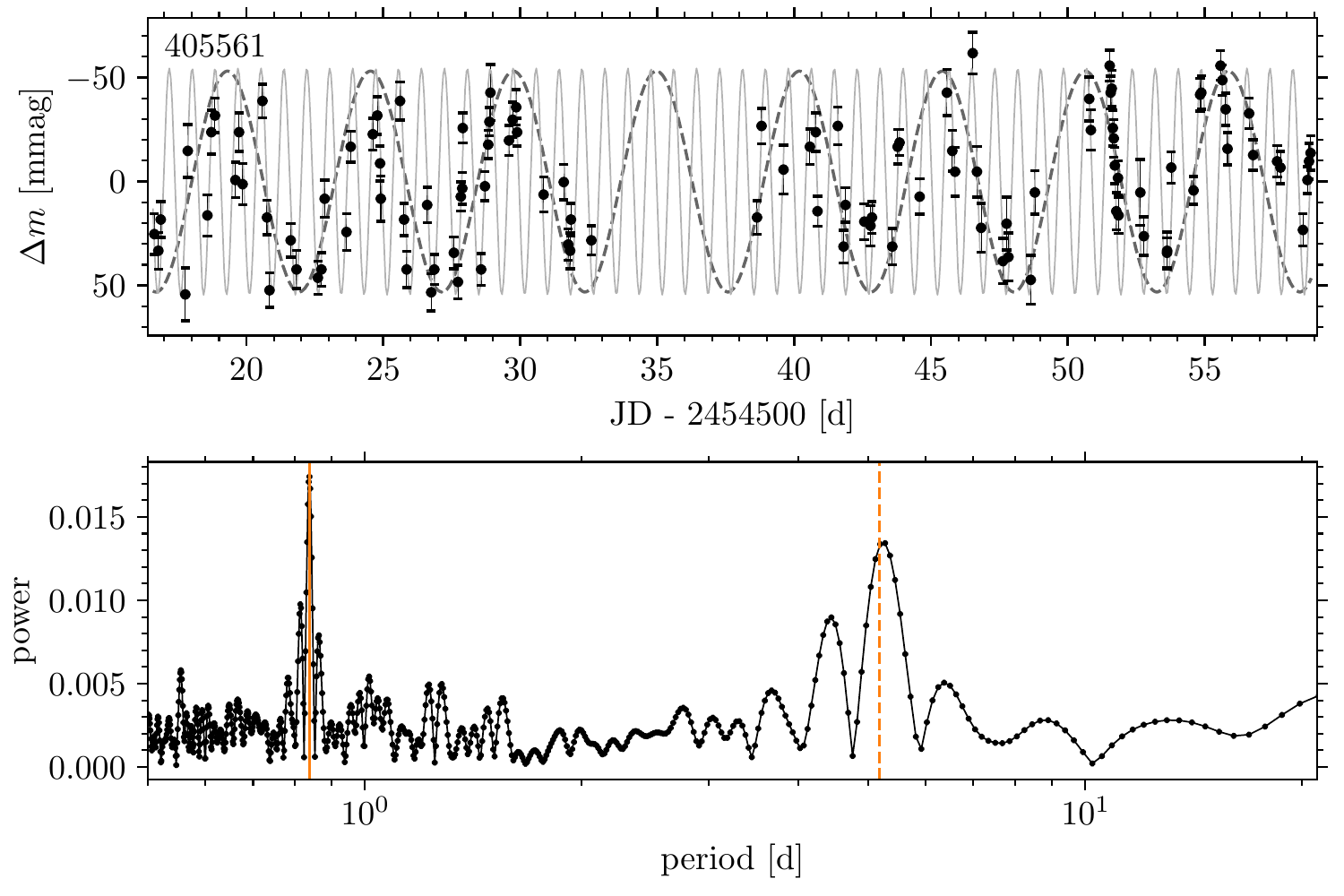}
	\caption{\emph{Top:} Example of a light curve with a strong alias. To guide the eye, sine waves for both the preferred period ($P_\mathrm{rot}=0.8385$\,d, solid line) and the alias period ($P_\mathrm{alias}=5.18$\,d, dashed line) are added to the data (black dots). The shorter period fits the data more closely as is shown by the fact that the simple sine connects nearly all data points. \emph{Bottom:} Corresponding \textsc{clean} periodogram with both periods marked with vertical lines. The alias period is not located exactly at the peak in the periodogram due to the coarser resolution at longer periods.}
	\label{fig:alc}
\end{figure}

In cases where possible alias periods are present, we carefully double-checked the light curves and periodograms before assigning a preliminary period. 
An example of such a case is displayed in Fig.~\ref{fig:alc}. 
The periodograms picked up both the preferred rotation period ($P_\mathrm{rot}=0.839$\,d) and the alias period\footnote{$P_\mathrm{alias}=1/\left(1/P_\mathrm{rot}-1\right)$} ($P_\mathrm{alias}=5.18$\,d), with equal power. 
Indeed, during our initial manual classification, we 
also interpreted this light curve as having
a ${\sim}5.2$\,d period 
with a noisy light curve, 
based on the majority of our periodograms.
However, the \textsc{clean} periodogram (designed precisely for such cases) damps down the beat frequency caused by the observing cadence, and enabled us to identify the alias to assign the star what we believe is the true period of $0.839$d.
In fact, manual inspection shows that successive data points on a given night follow the short-period sinusoid without exception, in marked contrast to the habitual excursions of the data from the longer-period sinusoid.

While some of the final published periods are multiples of 1\,d,
even the \textsc{clean} periodogram confirms them with high significance. 
We cannot detect any signs of possible aliases despite thorough analyses of the periodograms. 
Consequently, we believe these periods to be real (and indeed expected), and include them in the final data set. 
Finally, it should be noted that the analysis in the companion stellar activity paper has additionally confirmed all rotation periods in the set considered therein, consisting of those stars for which we were able to obtain spectroscopy, and therefore chromospheric activity measurements.

\subsection{Further steps to final periods}
\label{sec:meanPer}

The above-determined period for each light curve is considered preliminary and imprecise because each of the applied algorithms provides a slightly different estimate for the underlying final period.
We therefore combine the results of all methods by seeking the most likely period near the agreed period in each periodogram, and take the mean value from the five different methods as our final period. 
Our uncertainty is taken to be the maximum difference of any of the initial periods from the mean period. 
As seen in F20, periods for the same star from different light curves agree to within this uncertainty estimate at a $2\sigma$-level.

As a result of our visual inspection, we classified the light curves into three classes. 
Stars of the primary class (`1' in the corresponding table) show a clear rotational signal, in both the light curve and the periodogram, which could in principle be found visually without any periodogram analysis. Stars of the secondary class (2) have a noisier light curve, a wide peak in the periodogram, or the periodogram contains secondary peaks with significant power.  
We additionally marked all stars which could potentially be affected by an alias with a `0' in the table.

Among the radial velocity cluster members with light curves (549 stars with 1158 light curves), we find \nperlc{} rotators. 
Additionally, we searched for periodicity among the \emph{Gaia}~DR2 proper motion members from F19 for which we have not obtained radial velocity data, and find 11 additional rotators among them.

There is an additional set of photometric periods beyond those that are obvious from their light curve variability alone.
These are stars for which we have 
obtained chromospheric activity measurements, and for which the well-defined relationship between stellar activity and rotation period (really $\textrm{Ro}$, the Rossby Number) has allowed us to predict a rotation period for the star from the corresponding activity measurement. 
Subsequently, we were able to identify the actual photometric rotation period in our data, typically within an interval of $\pm 1$\,d of the predicted period.
This procedure is fully explained in our companion publication (F21act), and enables us to uncover \nperact{} additional (photometric) rotation periods from our light curves.
This third class of rotation periods, identified with the help of their
chromospheric activity measurements, are marked with the numeral `3' in the accompanying table.

Hereafter, we use the full set of \npertotal{} rotation periods regardless of their origin.
However, we mark them with distinguishing symbols in the main colour-period diagram in Sect.~\ref{sec:CPD} 
so that the reader can appreciate that their distributions are essentially indistinguishable.
All stellar rotation periods, accompanied by the corresponding period uncertainties, the classification and other relevant information, are provided in an accompanying online Table. (See Table~\ref{tab:rotationperiods} for a specification of the data available.)

\begin{table}
	\caption{Description of the columns of the online Table with the measured rotation periods.}
	\label{tab:rotationperiods}

	\begin{tabular}{lll}
		\hline
		\hline
		Name & Unit & Description\\
		\hline
		ID & - & ID in this work and F19\\
		designation & - & ID from \emph{Gaia} DR2\\
		CLWH & - & ID from \protect\cite{clem}\\
		RA & deg & Right ascension from \protect\cite{clem}\\
		Dec & deg & Declination from \protect\cite{clem}\\
		Vmag & mag & $V$ magnitude from \protect\cite{clem} \\
		BV0 & mag & $(B-V)_0$ colour from \protect\cite{clem}$^1$\\
		VK0 & mag & $(V-K_s)_0$ colour with $K_s$ from 2MASS\\
		Prot & d & Rotation period\\
		dProt & d & Uncertainty on rotation period\\
		Amp & mag & Light curve amplitude \\
		Class & - & Classification of period [0, 1, 2, 3]$^2$\\
		Binary &- & Binary status (True/False)\\
		\hline
	\end{tabular}
\tablefoot{$^1$ Dereddened with $E_{(B-V)}=0.034$, $E_{(V-K_s)}=0.095$ (F19).\\
	$^2$ Classes are 0: possible alias, 1: first class, 2: algorithmic periods, 3: activity-informed periods.}
\end{table}

\subsection{Light curve amplitudes}
\label{sec:varamp}
A simple and direct measure of photospheric activity that can be obtained from the light curves is the amplitude of variability.
This provides an indication of starspot size, convolved of course, with the star's rotational inclination angle, and the degree of asymmetry of starspot distribution with longitude.
For completeness, we also provide the variability amplitudes of the rotators in Table~\ref{tab:rotationperiods}.
However, the discussion of the variability amplitudes themselves, and their relationship with other activity indicators is 
carried out in the companion paper (Sect. 4 and Fig.~2 therein). 
In brief, as with other stellar activity indicators, the variability amplitudes follow a distinct correlation with the Rossby number 
(defined as the ratio between the rotation period and the convective turn-over time). 
Additionally, 
from the comparison with NGC\,2516, we show that variability amplitudes of the slow rotators decrease with age due in concert with the stellar spin-down and the accompanying decrease in magnetic field strength.

\subsection{Rotators in the colour-magnitude digram}

\begin{figure*}
	\includegraphics[width=\textwidth]{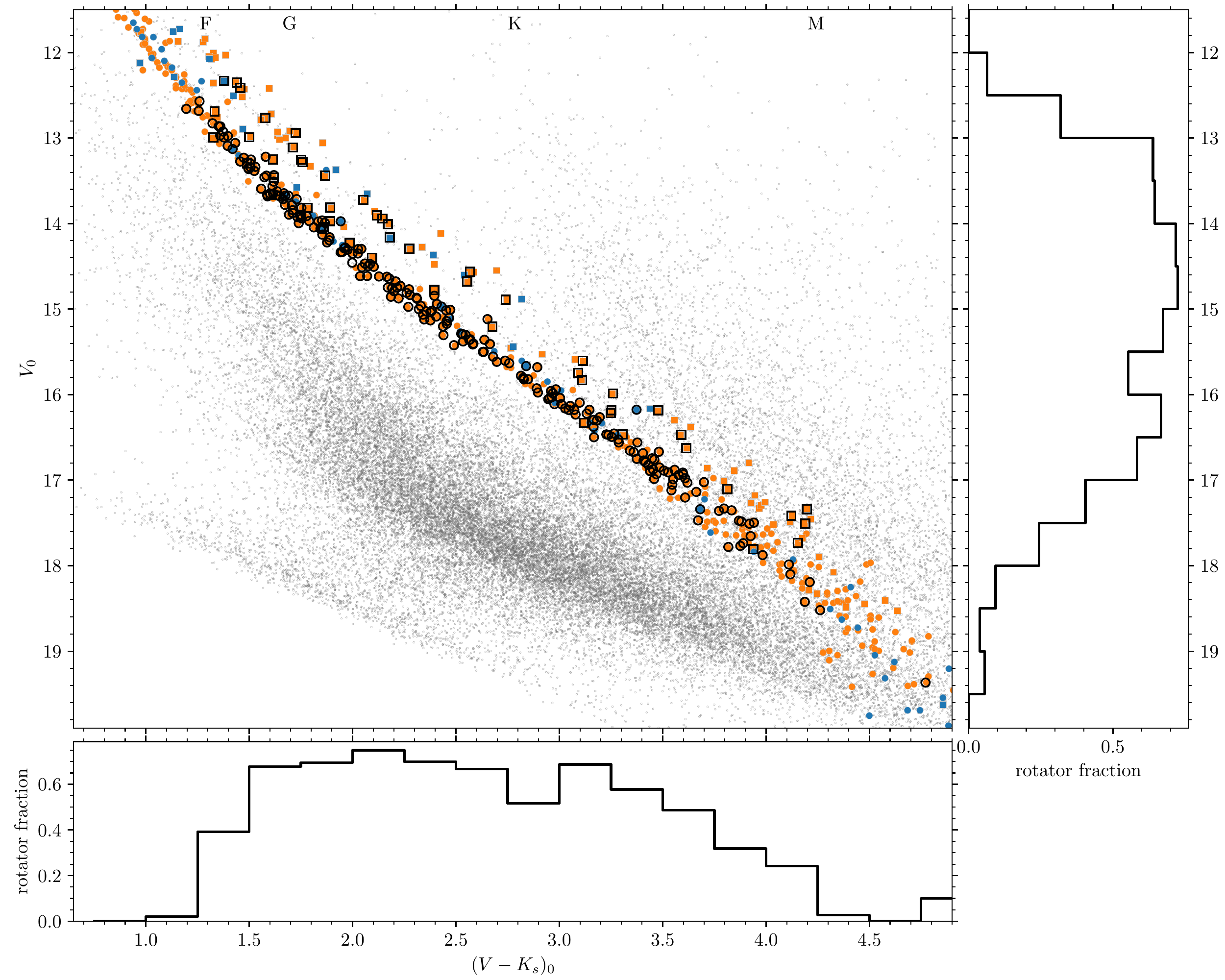}
	\caption{\emph{Left:} Colour-magnitude diagram of NGC\,3532 highlighting the identified rotators (black outline). Rotation periods have been measured for the majority (detection fraction > 50\,\%) in the G2 to K8 spectral type range. The colour scheme is the same as in Fig.~\ref{fig:memCMD}, with radial velocity members of NGC\,3532 in orange and additional proper motion members in blue. Likely cluster binaries are marked with squares. \emph{Right} and \emph{bottom:} Histograms showing the fraction of periodic rotators identified among the members. The decline for $(V-K_s)_0 > 3.2$ and $V_0 > 16.5$ arises from the decreasing sensitivity for fainter stars.}
	\label{fig:CMD}
\end{figure*}

To sum up our rotation period derivations for NGC\,3532, we mark all identified stars with derived rotation periods in the colour-magnitude diagram (Fig.~\ref{fig:CMD}). 
The rotators (including both those obtained only from the photometry, and those obtained from the photometry, but with the help of the activity information) 
can be seen to be distributed across the whole cool star region of the cluster sequence ranging from spectral type late-F to mid-M. 
The histogram in the right panel of Fig.~\ref{fig:CMD} shows a clear peak for stars around $V\approx 14.5$, corresponding to early K\,dwarfs in NGC\,3532 ($(V-K_s)\approx2$). 
Here, we detect rotation periods for 87\,\% of the NGC\,3532 members. 
Across the whole magnitude range $V=13-17$, we are able to find rotation periods for more than half of the member stars. 
For fainter and redder stars the detection fraction decreases strongly as a result of decreasing sensitivity, so that stars with $V\ge17$ are only occasionally identified as rotators.

\section{Colour-period diagram for NGC\,3532}
\label{sec:CPD}

Because the rotation periods of open cluster stars are now well-known to have a strong dependence on stellar mass, it is particularly useful to discuss the measured periods for the NGC\,3532 stars in concert with their photometric colours, a precisely measured proxy for their masses (or spectral types).
This section therefore discusses the NGC\,3532 colour-period diagram (CPD) that our rotation periods allow us to construct.

\begin{figure*}
	\includegraphics[width=\textwidth]{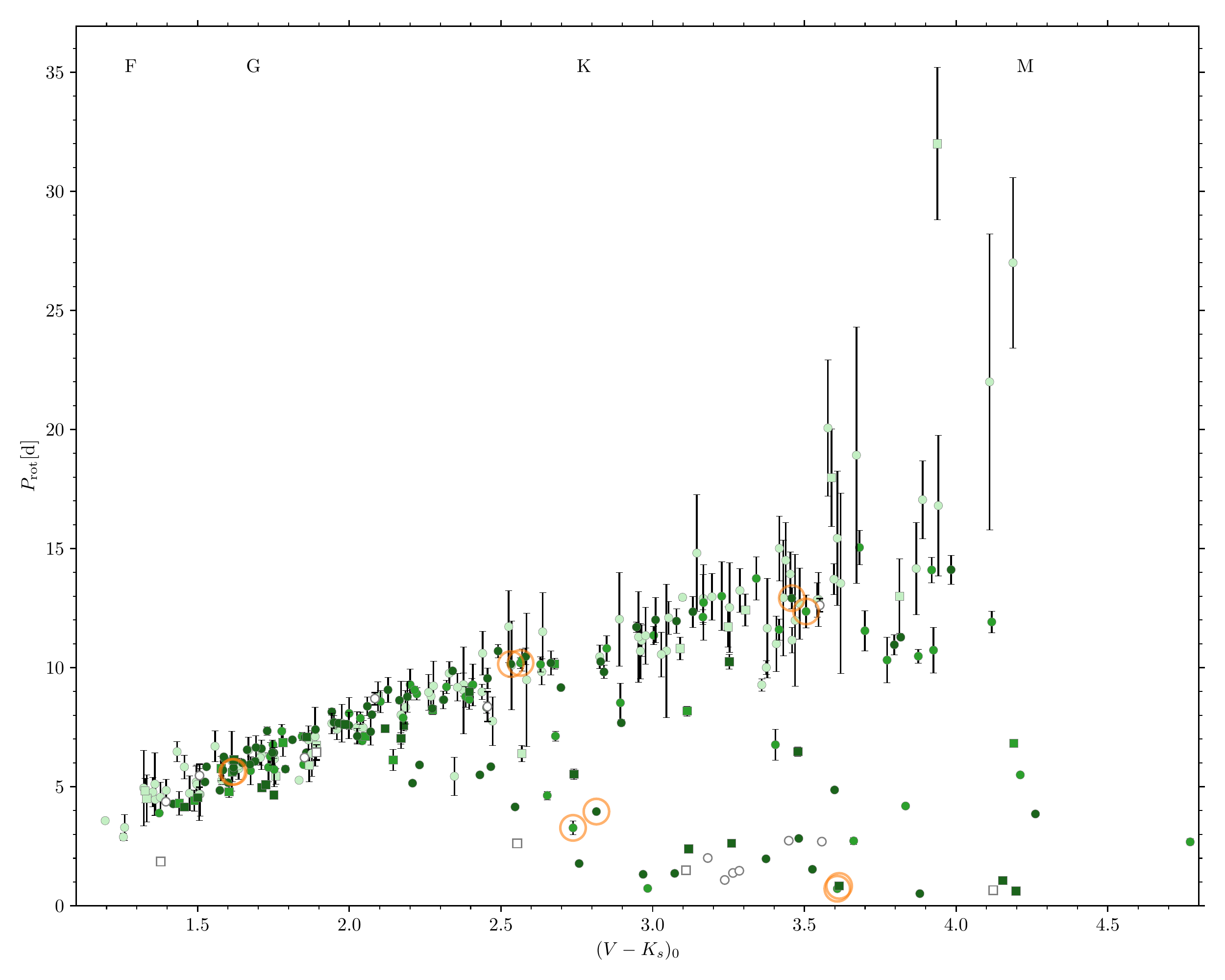}
	\caption{Colour-period diagram for members of NGC\,3532 with rotation periods.
	A well-populated slow rotator sequence is visible, as well an evolved fast rotator sequence. The continuation of the slow rotators into the M\,dwarf regime, the `extended slow rotator sequence' is also apparent. The paucity of warmer fast rotators relative to younger open clusters is remarkable, and emphasises the mass dependence of the transition from fast to slow rotation. The best photometric rotation periods are marked in dark green and less evident (algorithmic) periods in lighter green. The lightest green symbols indicate the activity-informed photometric rotation periods from the companion paper \citep{N3532act}. The few periods with strong aliases are marked with unfilled symbols. Circles indicate likely single members and squares binaries. We show example light curves in Fig.~\ref{appf:LCs} for the stars encircled in orange. The bluest datum consists of two stars with nearly identical positions in the CPD, so that only one circle is visible. The period uncertainties are only visible when they exceed the symbol size.}
	\label{fig:CPD}
\end{figure*}

This CPD for NGC\,3532, using the final set of \npertotal{} rotation periods, is displayed in Fig.~\ref{fig:CPD} using the $(V-K_s)_0$ intrinsic colour\footnote{With $V$ from \cite{clem} and $K_s$ from 2MASS \citep{2006AJ....131.1163S}.} as the independent variable.
(CPDs in other useful colours will also be provided below when relevant.)
Rotation period uncertainties are only visible when they exceed the sizes of the corresponding symbols, each explained in the figure caption.
The absence of outliers, especially in the upper left region of the CPD is notable\footnote{We ascribe this both to better sampling of the light curves, resulting in better rotation period derivation, and also the availability of better membership information, as compared with prior work, including earlier work on the same cluster in the otherwise unpublished PhD thesis work of \cite{1997PhDT.........7B}.}.

\subsection{Specific regions of the CPD}

The CPD of NGC\,3532 features a very prominent slow rotator sequence that stretches diagonally from ${\sim} 3$\,d for late F-type stars to beyond $15$\,d for early M\,dwarfs.
The highest-mass rotator in this range of spectral types has $M {\sim} 1.3\,M_\sun$ and the least massive $M {\sim} 0.5\,M_\sun$\footnote{Even lower mass stars are present, and will be discussed separately below.}. 
The sequence is remarkably well-populated and linear in this colour, as compared with open clusters of ZAMS age (discussed further below). 
The blue end of the slow rotator sequence begins at $P_\mathrm{rot} {\sim} 3$\,d. 
From space-based photometry of younger (e.g. NGC\,2516; \citealt{2020ApJ...903...99H}) and
older clusters (e.g. NGC 6811; \citealt{2015Natur.517..589M}, \citealt{2019ApJ...879...49C}) we know that stars blueward of our detections are likely to have even shorter rotation periods as they approach the break in the \cite{1967ApJ...150..551K} curve of rotational velocity against colour.
Our data are not sensitive enough to locate this break point in photometric colour,  well-known to be where the onset of surface convection zones among cooler stars trigger stellar spin-down via magnetised winds.
At the red end of the slow rotator sequence, 
we find an `extended slow rotator sequence', with early M-dwarfs rotating as slowly as 32\,d.
This feature appears to be the older counterpart of the extended slow rotator sequence (ESR sequence) that we identified in our prior work (F20) on the open cluster NGC\,2516.

We also find a sizeable number of fast rotators that clearly form a different population as compared with the slow rotators.
These are present among the cooler stars of our sample, and are essentially absent for stars bluer than $(V-K_s)_0 < 2.1$. 
The fastest rotators ($P_\mathrm{rot} \lesssim 3$\,d) of this group appear to delineate a flat sequence of roughly constant rotation period for $(V-K_s)_0 > 3$.
Their earlier counterparts with ($(V-K_s)_0 \approx 2-3$) have evolved noticeably upwards from their initially shorter rotation periods (as in the ZAMS clusters NGC\,2516 and Pleiades), and appear to have formed an elevated sequence of rotating stars in transition from fast to slow rotation. 
For even warmer stars neither the fast rotators nor their evolved counterparts are observed. 
In fact, as we show in the companion paper on the activity of NGC\,3532 stars, such stars do not exist in NGC\,3532, having all evolved into slow rotators.

The slow rotator sequence of NGC\,3532 appears to be largely straight, with rotation period increasing linearly with $(V-K_s)_0$ colour. 
Those of significantly younger or older open clusters feature more strongly-curved sequences. 
This linear appearance appears to be an intrinsic feature of open clusters near the age of NGC\,3532, as we shall see below in Sect.~\ref{sec:OCcompare}, where NGC\,3532 is compared with two open clusters (M\,34 and M\,48) that are immediately adjacent in age.
We also show below, in Sect.~\ref{sec:models}, that the curved slow rotator sequence of the younger NGC\,2516 open cluster in fact straightens out when it is evolved forward to the age of NGC\,3532 using evolutionary angular momentum models. 
Consequently, the change of shape of the slow rotator sequence appears to be a natural result of the mass-dependent angular momentum evolution, one which spins down the later-type stars on the slow rotator sequence much stronger than the warmer stars in the considered time frame.

Among the lower-mass late-K and early-M dwarfs ($(V-K_s)_0>3.5$), the membership information available is limited, and as a result, this region of the CPD is more sparsely populated. 
This particularly impacts the expected, but only barely sampled, wide distribution of rotation periods among early-M dwarfs (with $0.5\,\mathrm{d} \lesssim P_\mathrm{rot} \lesssim 10\,\mathrm{d}$) and the fast rotating tail which extends to the lowest-mass stars. 
Two additional factors are at work:
Firstly, the signal-to-noise ratio of the light curves is low for these faint stars, hindering unambiguous determination of the short rotation periods. 
Secondly, because the stellar activity of these stars is saturated,
we are unable to derive activity-informed rotation periods for such stars in the companion paper. 
Yet from the colour-activity diagram, we know that such stars are present in NGC\,3532.

The CPD has a sparsely populated gap in the vicinity of $(V-K_s)_0=2.75$. 
This is a consequence of the usage of $(V-K_s)_0$ colour, with a large extent in colour space representing only a small mass range, and hence fewer stars fall into that region in the CPD. (The CMD in Fig.~\ref{fig:CMD} also shows a similar feature.).
We find one obvious outlier below the slow rotator sequence among the bluest stars. 
We would not normally expect any stars in this region because such stars tend to evolve onto the slow rotator sequence. 
However, this rotation period is a possibly-aliased period of comparatively low quality. 
The membership of this star is also based only on \emph{Gaia} data, without radial velocity confirmation. 
The star could also potentially be in a tidally locked binary system, as is the case for one of the equivalent rotators in NGC\,2516. 
In conclusion, we cannot find a reason to disregard this rotator, and retain it as is for this work.

\subsection{Bifurcation of the slow rotator sequence}
\label{sec:bifurcation}

\begin{figure}
	\includegraphics[width=\columnwidth]{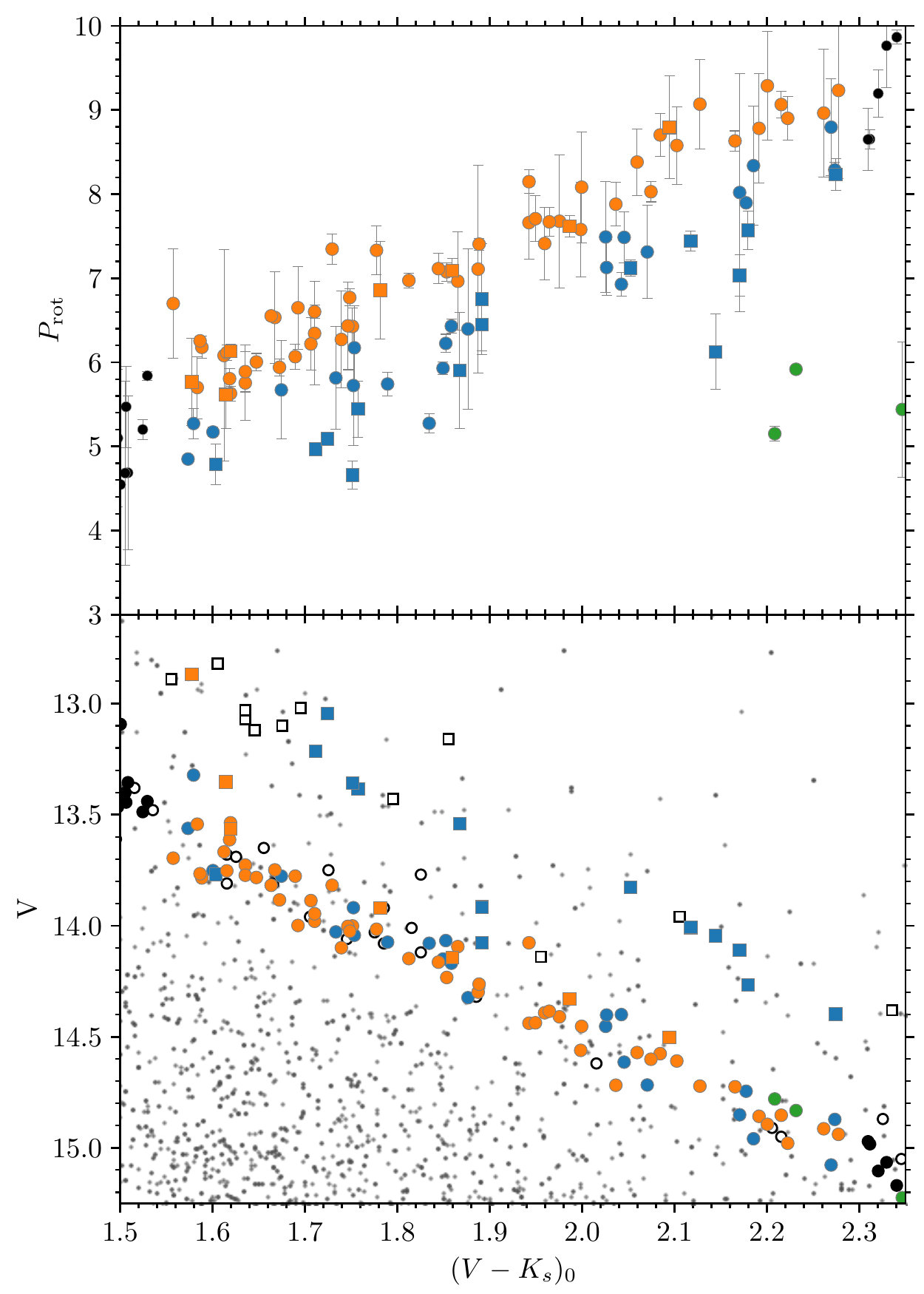}
	\caption{\emph{Top}: Zoom-in on the region of the colour-period diagram in which the bifurcation of the slow rotator sequence is observed. The longer-period sequence (S-slow rotators) is shown in orange and the shorter-period sequence (F-slow rotators) in blue. Additional slow rotators outside the region of interest are marked in black, and fast rotators in green. Squares indicate binaries.
		\emph{Bottom}: Corresponding colour-magnitude diagram (same symbols as above) with additional cluster members without determined rotation periods marked with open symbols. The equal-mass binary sequence is almost exclusively populated by the F-slow rotators, indicating a connection between faster rotation and binarity.}
	\label{fig:bisection}
\end{figure}

We now discuss a curious feature of the slow rotator sequence that has not, to our knowledge, been seen in any prior open cluster.
The slow rotator sequence itself (in the range $(V-K_s)_0=1.6-2.3$) shows a clear separation into an upper (slower = S) and a lower (faster = F) sequence, particularly when the activity-assisted rotation periods with larger errors) are ignored. 
The effect remains observable when the noisier activity-assisted periods are added, even though the larger period uncertainties of those stars introduce an additional scatter
(Fig.~\ref{fig:CPD}).

Such a spread in the slow rotator sequence cannot be attributed to differential reddening displacing a subset of stars in colour from the true sequence. 
Given the low reddening ($E_{(B-V)} = 0.034$\,mag, F19) towards NGC\,3532, such an effect is impossible. 
Colour effects could potentially also be introduced through inaccurate photometry. 
However, when we plotted our CPD with different colour combinations -- including $(G_\mathrm{BP}-G_\mathrm{RP})$ from \emph{Gaia}~DR2 -- the gap running through the middle of the slow rotator sequence remains, and appears to be independent of the chosen colour combination.
We also verified the correctness of our periods. 
Having re-examined the light curves and periodograms to exclude any possibly falsely-assigned periods, we conclude that all periods displayed are genuine.

We also checked whether the two sequences could belong to two different stellar populations. 
The results are negative, with both groups apparently spread out evenly in proper motion, radial velocity, parallax, and sky position, disallowing recourse to a merger scenario for the NGC\,3532 cluster, for instance.

To display this effectively, we colour-code the slower- and faster rotational branches of the slow rotator sequence (hereafter S-slow rotators and F-slow rotators, respectively) as indicated in the CPD (Fig.~\ref{fig:bisection}, upper panel) and also plot them separately in the colour-magnitude diagram (Fig.~\ref{fig:bisection}, lower panel). We see that 
the equal-mass binary sequence in the CMD is essentially composed of only the F-slow rotators. 
These stars have the same colours as their individual components in the CMD.
Consequently, no colour shift is possible in the CPD either. 
The few photometric binaries between the single star main sequence and the equal-mass binary sequence\footnote{These are expected to be displaced slightly redward.} 
also tend to belong to the F-slow rotators, while the astrometric binaries (orange squares) on the cluster sequence belong to the S-slow rotators. 
We note that not all F-slow stars are demonstrably binaries. 
Among the F-slow rotators, the binary fraction is $37.8\pm10.1\,\%$, 
somewhat elevated with respect to the overall binary fraction in this colour range, which is $27.2\pm4.7\,\%$. 
Despite the higher binary fraction, we also find apparently single stars (based on decade-baseline radial velocity time-series, F19, and astrometry, Sect.~\ref{sec:binaries}) among  the F-slow rotators, a fact that suggests that the bifurcation cannot be attributed to tidal effects alone.

We note that a statistically significant rotational offset between the overall rotation period distributions of single and primary stars in close binaries beyond the reach of tides has also been noted by \cite{2007ApJ...665L.155M}.
Furthermore, we note that the binary fraction among the fast rotators of NGC\,3532 is also significantly elevated.
For the fast- and evolved fast rotators, we find $32.4\pm9.4\,\%$ to be binaries. We can compare this number to the model of \cite{2020ApJ...901...49L}, who estimate the binary fraction in different mass-regimes in NGC\,3532. 
They find a binary fraction\footnote{They include all binaries, in particular non-equal-mass binaries on the photometric single stars sequence.} of $19.4\pm2.2\,\%$ for lower-mass stars with $G<15.06$. 
This magnitude cut coincides with the bluest stars on the evolved fast rotator sequence.
Since this number is smaller than the observed fraction, we conclude that the fast rotator sequence of NGC\,3532 is enhanced in binaries as compared with the overall population, although it also includes many stars with no trace of binarity whatsoever.

The connection between fast rotation and binarity is not necessarily a tidal interaction, but could be a natural result of star formation. 
The binarity could influence the stellar rotation indirectly through the dissipation of the proto-planetary disc \citep{1994ApJ...421..651A,1994ARA&A..32..465M}. \cite{2009ApJ...696L..84C} have found short disc lifetimes in binary systems and \cite{2019A&A...627A..97M} has observed that young binary stars generally rotate faster. 
All together, one finds that the majority of binary stars rotate faster than average at young ages.

Continuing this chain of reasoning, we conclude that the F-slow rotators were initially fast rotators because: (1) they rotate faster than the S-slow rotators; (2) they have a larger binarity fraction, and (3) the initial fast rotators have evolved to slow rotators while preserving their binarity fraction during the transition. 
The stars of this group rotate faster because they have not yet settled onto the asymptotic slow rotator sequence. 
Such an effect could be the result of stellar spin-down being strongest during the crossing of the rotational gap \citep{2010ApJ...722..222B}, and weakening near the slow rotator sequence. 
When arriving on the slow rotator sequence, the braking is only marginally stronger than for the initial slow rotators which spin down Skumanich-like. 
Hence, the former fast rotators require some time to finally settle onto the sequence.

Our observations suggest that the influence on rotation beyond tidal interactions is apparently restricted to binaries of intermediate separations which are {both unresolved} and have a near-equal mass ratio (i.e. photometric binaries). 
Pure astrometric binaries are less likely to be found on the F-slow branch. 
Such stars could either be very wide binaries, with both components photometrically resolved, or have a low mass ratio, placing them on the single star main sequence. 
Disentangling this effect will clearly require more detailed work.

In conclusion, we can observe the transition from fast to slow rotation not only when the stars are spinning down through the rotational gap but also post that extremely rapid phase. 
Unfortunately, NGC\,3532 is unique in its age, richness, and proximity. 
Hence it is currently not possible to observationally test\footnote{As seen in Sect.~\ref{sec:models}, theoretical angular momentum models are not yet advanced enough to test such small effects through simulations.} whether the bifurcation exists in other open clusters of similar age, or to specify how long it might persist.

\section{Comparison with rotational isochrones}
\label{sec:modelgyrochrones}
\begin{figure}
	\includegraphics[width=\columnwidth]{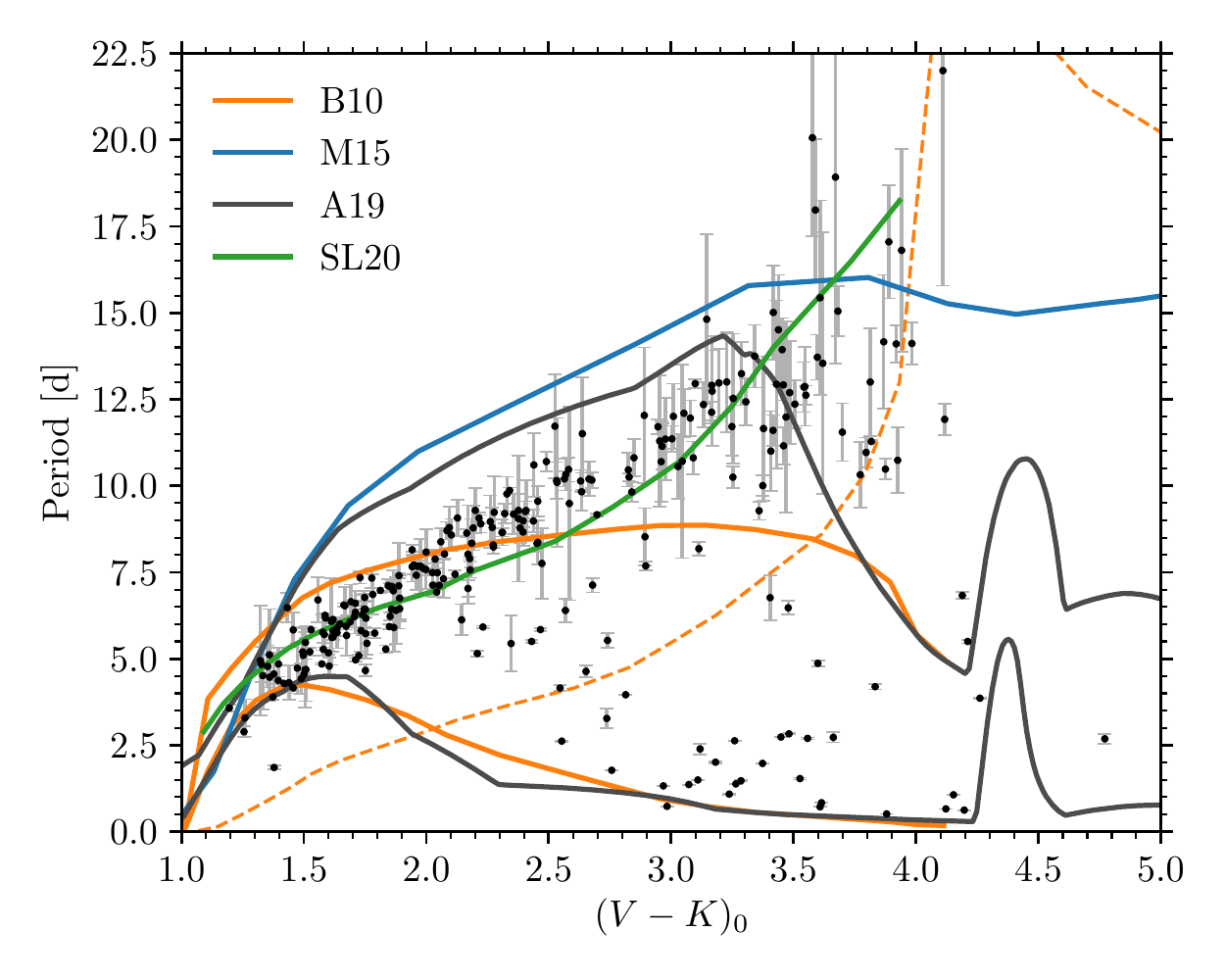}
	\caption{Comparison between the NGC\,3532 rotation periods and 300\,Myr stellar rotation models. In orange, we show both the slow and fast rotator model from \cite{2010ApJ...722..222B}. The dashed line indicates the position of the rotational gap from B10. The asymptotic spin rate from \cite{2015ApJ...799L..23M} for 300\,Myr is shown in blue and the isochrones from \cite{2019A&A...631A..77A} for both the fast and slow rotators are given in grey, while the slow rotator isochrone from \cite{2020A&A...636A..76S} is shown in green.}
	\label{fig:models}
\end{figure}

A number of models of stellar rotation have provided `rotational isochrones', enabling a direct comparison between the predictions of the model and the measured rotation periods across a wide range of cool star masses.
These isochrones in the (colour/mass)-period space can be placed directly onto the data to enable, among other things, an estimation of the rotational age of the open cluster with respect to the model in question. Here, we examine four recent models, all of which were compared in detail in connection with our prior work (F20) in NGC\,2516. 
Below, we only provide the briefest description of each model.
In order to intercompare the models on the same basis, we use the same age (300\,Myr) as the relevant comparison age for all models.
These are displayed 
in comparison with one another and with the measured rotation period data for NGC\,3532 in Fig.~\ref{fig:models}.

\subsection{\cite{2010ApJ...722..222B}}
The oldest model in our comparison (\citealt{2010ApJ...722..222B}, hereafter B10) has a mathematically simple symmetric form in the period derivative, $\mathrm{d}P/\mathrm{d}t$, and provides isochrones which depend only on the convective turn-over timescale $\tau_c$ 
which serves as the underlying independent variable for the stellar mass, and for which we use the values provided in \cite{2010ApJ...721..675B}.
(The table therein itself provides the $(V-K)$ colours required for the comparison.)

This model is plotted using orange lines in Fig.~\ref{fig:models}, the lower one corresponding to an initial period $P_0 = 0.1$\,d, and the upper one to $P_0 = 3.4$\,d, uniform for stars of all masses.
The dashed orange line indicates the boundary between fast and slow rotators in the B10 model.
We see that the location of this boundary line is plausible, in the sense that, as predicted by the model, the density of rotation period measurements is actually somewhat lower in its vicinity.
However, the line clips the measured slow rotator sequence among the M\,dwarfs, suggesting an issue with either the rotation period prediction or with the predicted stellar colours.
The prediction for the fast rotators is also good for stars of solar mass and warmer, as well as for the late-K and early\,M stars.
However, stars of late-G and early-K spectral type ($1.8 < (V-K_0) < 2.6$) 
clearly spin down faster than the model.
The prediction of the B10 model for the slow rotators is clearly above the NGC\,3532 data blueward of $(V-K)_0 = 2.2$, and below for redder colours.
A large part of this mismatch can be traced to the uniform (and therefore unrealistic) initial period $P_0 = 3.4$\,d that was chosen to generate this isochrone.
In Sect.~\ref{sec:models} below, we show that using the measured NGC\,2516 periods at 150\,Myr as a starting condition goes much of the way towards fixing this problem.

\subsection{\cite{2015ApJ...799L..23M}}
The \cite{2015ApJ...799L..23M} model (hereafter M15) also uses the convective turnover timescale, $\tau_c$, in cool stars via the Rossby Number, $Ro= P/\tau_c$.
The problem is formulated in terms of a spin-down torque that is expressed as a function of the mass, radius, and magnetic field of the star, with the magnetic field itself parametrised as a function of the Rossby number, mass, and radius.
The result produces a mass dependence of the spin-down timescales in both the fast and slow rotator regimes that is similar to that in the B10 model (see Fig.~1 in \cite{2015ApJ...799L..23M}), albeit with additional parameters.
This spin-down formulation is then applied to stellar evolution models from \cite{1998A&A...337..403B} to obtain the rotational evolution results.
The asymptotic spin rate for this model for $300$\,Myr is shown in Fig.~\ref{fig:models} with the blue curve. 

As can be seen in Fig.~\ref{fig:models}, the 300\,Myr model appears at face value to have over-estimated the rotation periods over nearly the whole mass-range. 
It is only among the F-type stars that it approaches the slow rotator sequence, but it does so with too steep a slope as compared to the NGC\,3532 observations. 
However, the isochrone from this model promisingly runs roughly parallel to the data redward of $(V-K_s)_0\approx 1.6$, suggesting that parameter changes (such as age) could potentially permit closer reproduction of the data.
The fast rotators in NGC\,3532 cannot be addressed by the asymptotic spin rate formalism.
We show a more detailed comparison below that includes the fast rotators,  and also relaxes the age constraint of 300\,Myr.

\subsection{\cite{2019A&A...631A..77A}}
\cite{2019A&A...631A..77A} (A19) have integrated the M15 angular momentum model into the Geneva/Montpellier stellar models, while also updating some parameters and tuning the initial conditions to reproduce rotation period distributions of various open clusters. 
Consequently, the prediction of their model is similar to that of M15 in the slow rotator regime, where it also over-predicts the rotation periods with the 300\,Myr model. 
In contrast, the fast rotator isochrone that is included in A19 predicts faster rotation than is observed in NGC\,3532 over the G-K mass range. 
As a result, the predicted dispersion in rotation period is far greater than that actually observed in this mass range.
We discuss implications of this mismatch in Sect.~\ref{sec:fastevolution} below.

\subsection{\cite{2020A&A...636A..76S}}
The final isochrone considered is that from \cite{2020A&A...636A..76S} (SL20), one formulated only for the slow rotators in open clusters.
The key feature of this model is the (re-)introduction of the two-zone model in which the radiative core rotates initially independently of the outer convective envelope\footnote{This concept was initially modelled by \cite{1991ApJ...376..204M}.}. 
The associated coupling is superimposed on the spin-down formulation from the B10 model, requiring one additional parameter that describes the power-law dependence of the coupling timescale on stellar mass.
It allows the angular momentum reservoir hidden within the star to be tapped on that mass-dependent timescale, correspondingly allowing cool stars to avoid spinning down at key ages, as observed in older open clusters.
For instance, it is the model that most accurately describes the rotation period distribution in the 2.5\,Gyr-old Ru\,147 \citep{2020A&A...644A..16G}.

From the rotational coupling time-scale determined by \cite{2020A&A...636A..76S} (their Fig.~6) one finds that for a 300\,Myr-old open cluster the coupling is dominant for stars with $M_\star \gtrsim0.9\,M_\sun$ 
(equivalent to $(V-K_s)_0 \lesssim 1.9$).
The two-zone model should describe the rotation period distribution better than a model based on pure wind braking in this regime. 

Indeed, Fig.~\ref{fig:models} shows that in this colour range \cite{2020A&A...636A..76S} describe the slow rotator data adequately. 
However, for lower mass stars the rotation periods appear to be slightly underestimated in a mass-dependent way.
This means that the braking is not efficient enough for these stars. 
Yet for the lowest-mass stars in the model, it is in good agreement with the observations of the beginning at the extended slow rotator sequence. 
The fast rotators are not modelled by \cite{2020A&A...636A..76S}.

\subsection{Conclusions about rotational isochrones}

In summary, with respect to the slow rotator sequence, we find the two-zone model of \cite{2020A&A...636A..76S} to provide the closest match to the NGC\,3532 rotation period measurements. 
The M15 and A19 models significantly overestimate the rotation periods for the slow rotators in the G-K\,star regime, at least to the extent that $300$\,Myr models are used.
The B10 model, the earliest of the set, is inadequate for the slow rotators, probably because of the unrealistic uniform initial rotation period of $3.4$\,d chosen.
As regards the fast rotators, only B10 and A19 make readily testable predictions.
These are nearly identical, transitioning stars warmer than solar onto the slow sequence as observed, reproducing the lower envelope for warmer than G2 spectral type and cooler than mid-K. 
However, neither of them appears to spin stars down aggressively enough in the region between.
We conclude that while the currently available theoretical work is plausibly in the region of the rotation period data, detailed reproduction of the data will require further work, even for the NGC\,3532 cluster itself.
In the following, we directly employ spin-down relations in order to understand these discrepancies.

%
%

\section{Detailed comparison with other open clusters and evolutionary models}
\label{sec:detailedComp}

The comparison between the NGC\,3532 rotation period data and the available rotational isochrones being less than satisfactory, we now perform a direct comparison with rotation periods from two open clusters that most closely bracket NGC\,3532 in age.
We then take the rotation period distribution of the ZAMS open cluster NGC\,2516 that was studied by F20 (one with essentially the same rotation period sensitivity), 
and evolve it forward in time to examine the extent to which two rotational evolution models are able reproduce the NGC\,3532 data on a star-by-star basis, while relaxing certain assumptions and choices inherent in constructing rotational isochrones.
This procedure is a deeper (and more fair) comparison than the prior one, and permits an appreciation of how the rotational age of a cluster currently depends on the underlying spin-down model.

\subsection{Comparison with clusters bracketing NGC\,3532 in age}
\label{sec:OCcompare}

\begin{figure*}
	\includegraphics[width=\textwidth]{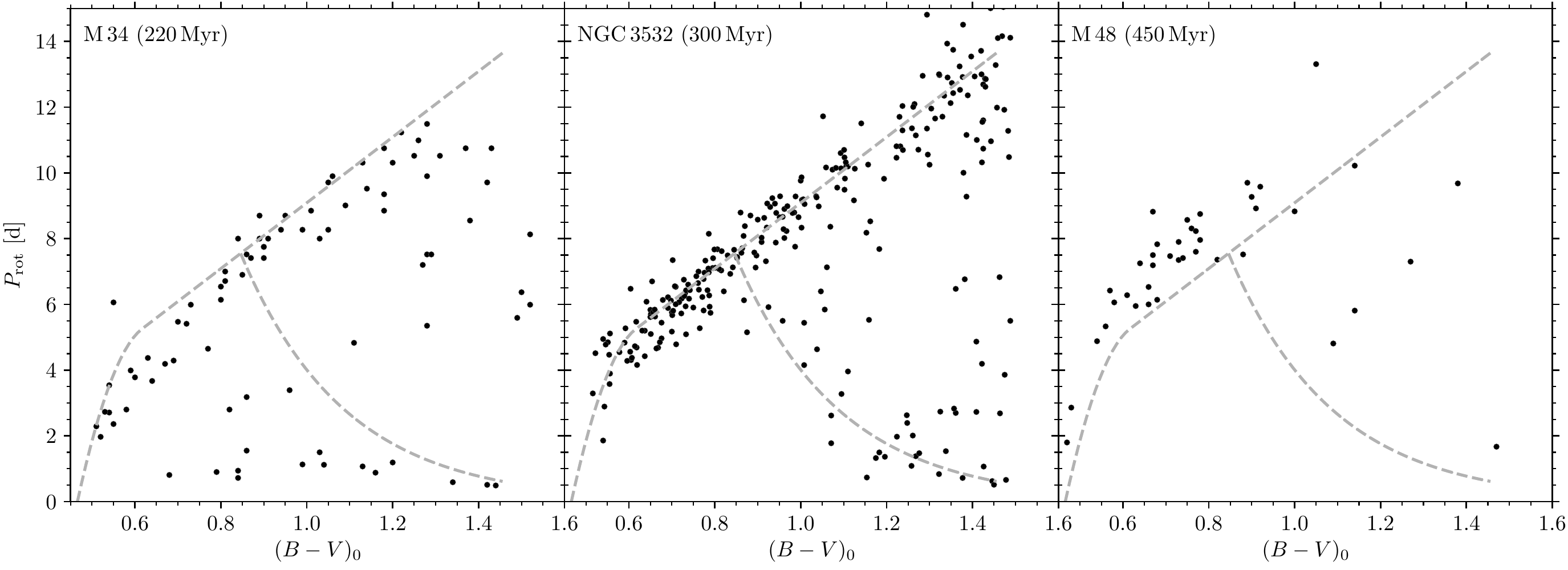}
	\caption{Colour-period diagrams for M\,34 \citep{2011ApJ...733..115M}, NGC\,3532 (this work), and M\,48 \citep{2015A&A...583A..73B} (from left to right) showing the period evolution in this $220-450$\,Myr age range. Fiducial lines for NGC\,3532 (dashed) are reproduced in the panels for M\,34 and M\,48. We see that the M\,34 rotation period distribution is mostly below the NGC\,3532 fiducials while the M\,48 distribution is above them. This informs us that M\,34 is younger than NGC\,3532 while M\,48 is older.}
	\label{fig:cOC}
\end{figure*}

The unique age of NGC\,3532 prevents us from directly comparing it to another coeval open cluster (as we were able to do in F20 for the case of NGC\,2516). 
In Fig.~\ref{fig:cOC}, we compare the rotation period distributions of NGC\,3532 (300\,Myr, centre panel) to the nearest younger cluster with measured rotation periods, M\,34 (220\,Myr, \citealt{2011ApJ...733..115M}, left panel), and to the nearest older cluster, M\,48 (450\,Myr, \citealt{2015A&A...583A..73B}, right panel). 
These are all post-ZAMS clusters, each stepping forward $\sim 100$\,Myr in age.
We wish to know (among other things) whether the rotational distributions are sufficiently different to be visually recognisable.
To enable a visual comparison, we distil the NGC\,3532 rotation period distribution into dashed lines that empirically represent the fast and slow rotator sequences in a similar manner to that in \cite{2003ApJ...586..464B}, and place that distilled empirical NGC\,3532 distribution onto the comparison cluster data in each panel of Fig.~\ref{fig:cOC}.

Using the lines of the distilled NGC\,3532 distribution, it is immediately apparent that stars on the slow rotator sequence of M\,34 are mostly below the lines, and thus rotate faster for a given mass (or colour) than stars in NGC\,3532. 
In contrast, the slow rotators in M\,48 are almost all located above the NGC\,3532 lines, and can be seen to rotate significantly slower than their counterparts in NGC\,3532. 
This age-ranked succession allows us to conclude immediately that the rotational age of NGC\,3532 is between those of M\,34 and M\,48, and must be in the vicinity of $\sim 300$\,Myr.
This is in qualitative agreement with the isochronal age of the cluster, assuming that those of the comparison clusters are substantially correct.

The evolution and dissipation of the fast rotator sequence, in contrast, is much greater than the small (if significant) evolution of the slow rotator sequences from M\,34 age to M\,48 age.
In M\,34, early K dwarfs ($(B-V)_0 \approx 0.8$) are still on this fast sequence. By the age of NGC\,3532, they have evolved from $P_\mathrm{rot} \approx 1$\,d to slow rotators with $P_\mathrm{rot} \approx 6$\,d. 
In NGC\,3532, stars rotating that fast can now only be found among the mid- to late-K and M\,stars. 
In M\,48, no such stars are observed, although this could potentially be an effect of the sampling of the light curves, and the relative insensitivity to late K and early M dwarfs in the study of \cite{2015A&A...583A..73B}. 
Nevertheless, the dramatic evolution of the fast rotators is potentially a sensitive age indicator which could be used to distinguish between open clusters of very similar isochronal ages.

We conclude that a ${\sim} 100$\,Myr age difference between young (below Hyades-aged) open clusters is large enough that the 
differences between their rotational distributions
can be appreciated by simple inspection, without the necessity for sophisticated analysis.

\subsection{Comparison with spin-down models}
\label{sec:models}

We now evolve the rotation period distribution of the ZAMS open cluster NGC\,2516 from F20 forward to mimic the measured NGC\,3532 distribution.
We have shown in F20 that the rotational distribution on the zero-age main sequence is universal 
in the sense that the distribution of one cluster is indistinguishable from those of other comparable ZAMS open clusters. 
Furthermore, both NGC\,2516 and NGC\,3532 were observed quasi-simultaneously during the same observing run, with the same instrumentation, observing baseline, near-similar cadence, and analysis techniques, ensuring that the rotation period sensitivities for both clusters are almost identical.
These features make the two datasets the best-comparable ones.

\begin{figure}
	\includegraphics[width=\columnwidth]{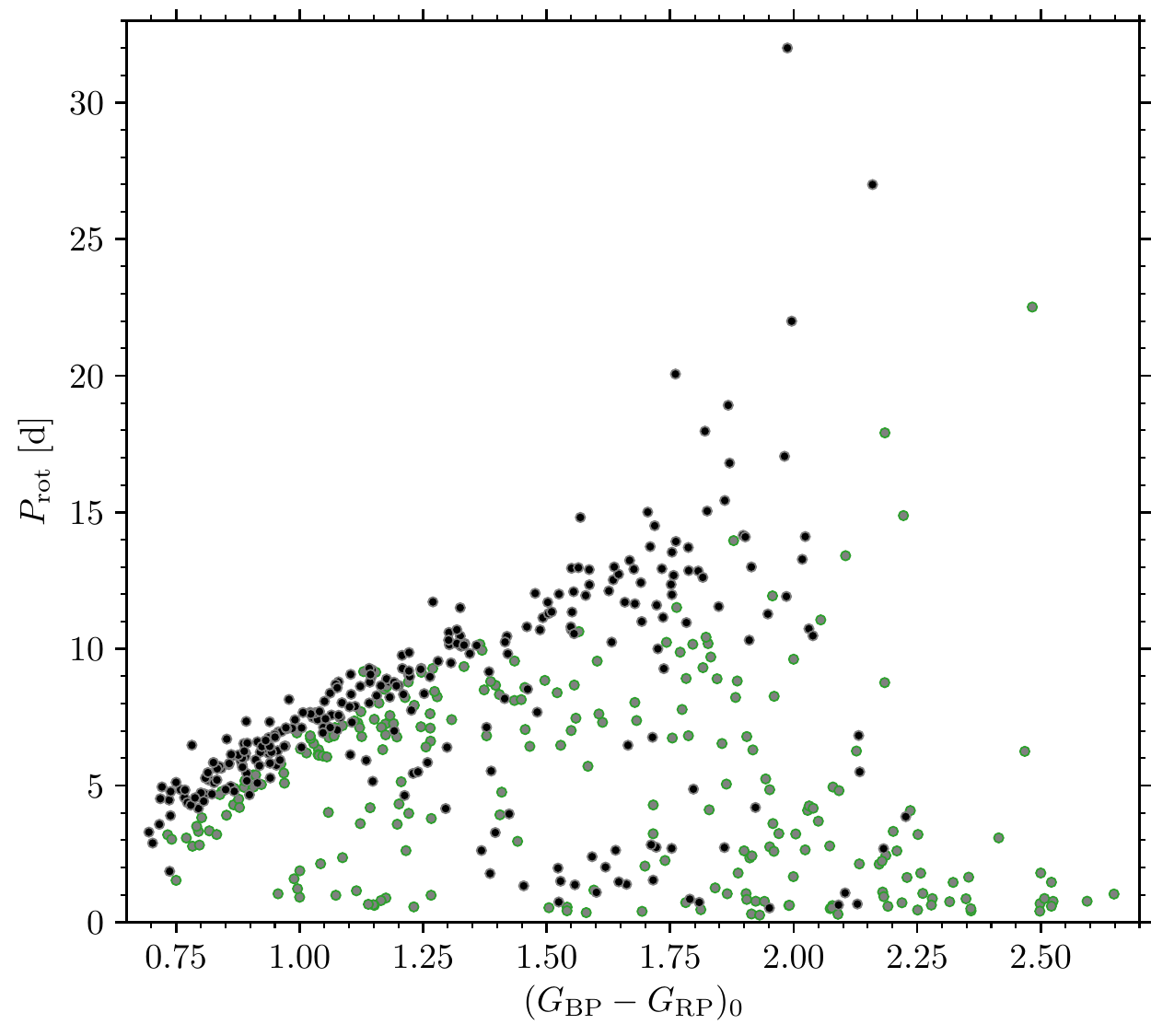}
	\caption{Colour-period diagram comparing NGC\,2516 (green, 150\,Myr, F20) with NGC\,3532 (black, 300\,Myr) based on \emph{Gaia}~DR2 $(G_\mathrm{BP}-G_\mathrm{RP})_0$ colour. The upward displacement of the slow rotator sequence of NGC\,3532 relative to NGC\,2516, as well as the dramatic evolution of its fast rotators onto the slow rotator sequence are visible well.}
	\label{fig:comp2516}
\end{figure}

The rotational distributions of both datasets are correspondingly shown
in the CPD in Fig.~\ref{fig:comp2516} 
using $(G_\mathrm{BP}-G_\mathrm{RP})_0$ colour \citep{2018A&A...616A...1G,2018A&A...616A...4E}. 
Both distributions can be observed to have the now-familiar cuneiform shape that characterises the CPDs of young open clusters, with well-defined slow rotator sequences, and no outliers in the upper left of the figure.
We see that in the interval between the ages of NGC\,2516 and NGC\,3532, the slow rotators have clearly moved upwards across every mass bin sampled. 
We also see that the fast rotators with $(G_\mathrm{BP}-G_\mathrm{RP})_0\le1.5$ have experienced very strong spin-down, 
while those that are redder display little, if any, spin-down.

The extended slow rotator sequence also appears to have 
spun down significantly, although the relatively small number of periods derived is a potential issue.
We concentrate below on the traditional slow and fast rotators and explore whether spin-down models can reproduce their evolution.

For this evolution, we apply the two spin-down prescriptions provided in \cite{2010ApJ...722..222B} (B10) and \cite{2015ApJ...799L..23M} (M15), both of  which can be implemented easily. 
Despite the similarity in their predictions, both models are calculated in significantly different ways.
While the B10 model is formulated solely in terms of the convective turn-over timescale (which then accounts for the mass dependence of the spin-down in that model), the M15 model requires the stellar parameters from evolutionary models. 
Although the latter is tuned to be used with \cite{1998A&A...337..403B}, a different stellar evolution model can in principle be used, and we correspondingly apply the updated models from \cite{2015A&A...577A..42B}\footnote{We also tested other stellar models, finding the final results to be the same within the uncertainties.}. 
For both models, we use the same convective turn-over time ($\tau_c$) as in the original works. 
Hence, the B10 model is calculated with \cite{2010ApJ...721..675B} and the M15 one with \cite{2011ApJ...741...54C}. 
(The two differ essentially by only a scaling constant.)

For simplicity, we assume in all calculations that an assigned mass or convective turn-over time is constant over time. 
Because most stars have already reached the main sequence by the age of our clusters, this assumption is justified. 
To evolve the periods, we applied to each individual measured period in NGC\,2516
Eq.~10 from B10:
\begin{equation}
	\frac{\mathrm{d}P_\mathrm{rot}}{\mathrm{d}t}=\left(\frac{k_I P_\mathrm{rot}}{\tau_c}+\frac{\tau_c}{k_C P_\mathrm{rot}}\right)^{-1}
\end{equation}
(with $k_I=452$\,Myr\,d$^{-1}$ and $k_C=0.646$\,d\,Myr$^{-1}$), exactly as in B10.

For M15, we apply Eq.~14 from M15:
\begin{equation}
	\frac{\mathrm{d}\Omega_\star}{\mathrm{d}t}=\frac{T}{I_\star}-\frac{\Omega_\star}{I_\star}\frac{\mathrm{d}I_\star}{\mathrm{d}{t}},
\end{equation}
where $I_\star$ is the stellar moment of inertia from a stellar model, $\Omega_\star$ is the angular rotation rate, and $T$ is the wind torque (Eqs. 6 and 7 from M15):
\begin{equation}
	T=\begin{cases}
		-T_0 \left(\frac{\tau_c}{\tau_{c,\sun}}\right)^2 \left(\frac{\Omega_\star}{\Omega_\sun}\right)^3 &\qquad(\text{unsaturated})\\
		-T_0 \chi^2 \left(\frac{\Omega_\star}{\Omega_\sun}\right) &\qquad(\text{saturated})
	\end{cases}
\end{equation}
with the Solar values of $\tau_{c,\sun}=12.9$\,d and $\Omega_\sun=2.6\cdot 10^{-6}$\,Hz. 
$\chi$ is the inverse solar-scaled Rossby number, with the desaturation-limit $\chi=10$ and
\begin{equation}
	T_0= 6.3 \cdot 10^{30}\,\mathrm{erg}\left(\frac{R_\star}{R_\sun}\right)^{3.1} \left(\frac{M_\star}{M_\sun}\right)^{0.5},
\end{equation}
as in \cite{2019ApJ...870L..27M}.
We use a simple forward Euler method and chose an increment of 1\,Myr.

We focus first on the evolution of the slow rotators to estimate gyro-ages 
(= rotational ages). 
Later, we consider the evolution of the fast rotators, and in particular the stars in transition from fast to slow rotation.

\subsubsection{Slow rotators and the Gyro-age of NGC\,3532}

\begin{figure}
	\includegraphics[width=\columnwidth]{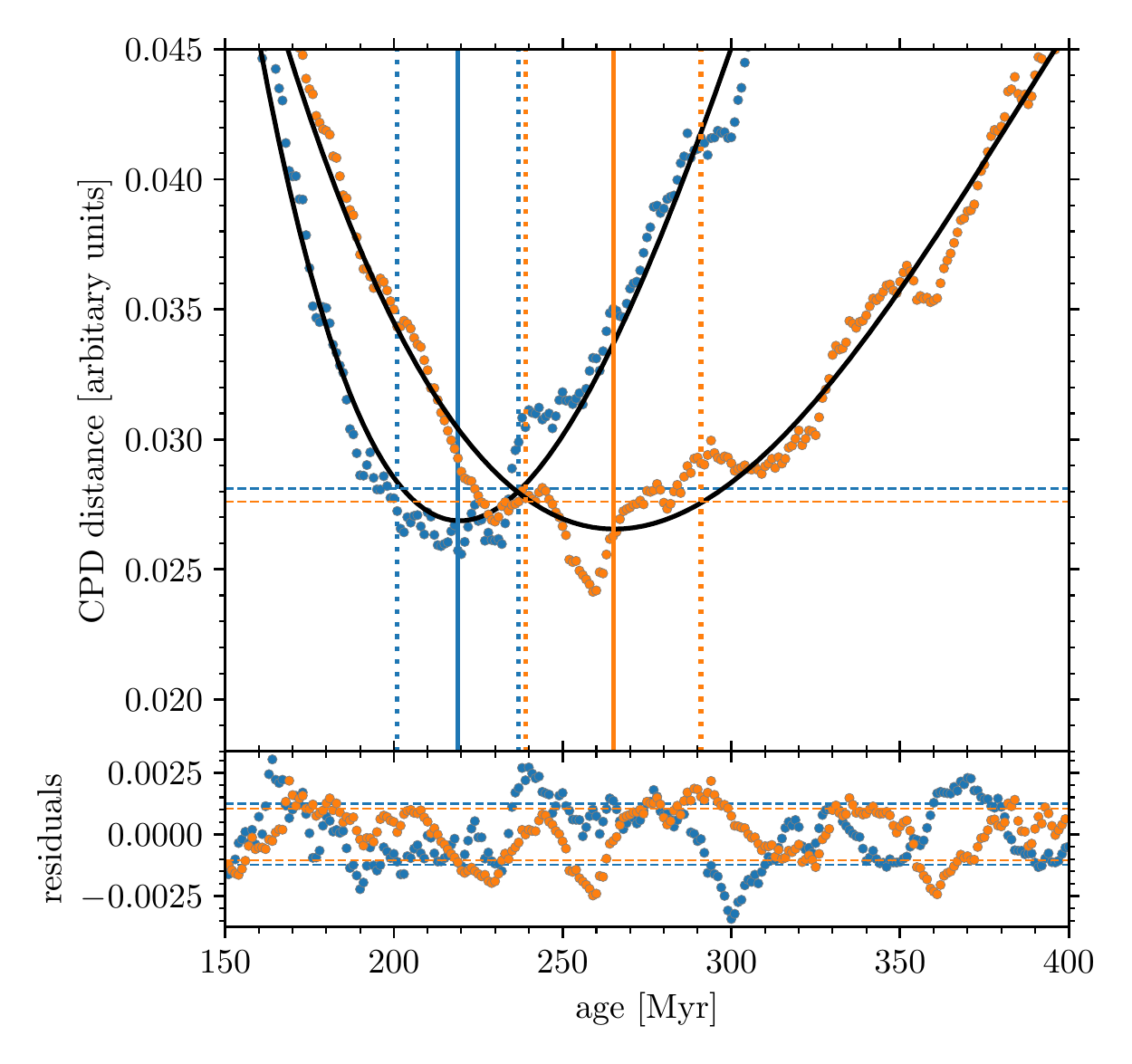}
	\caption{Rotational age for NGC\,3532 with respect to the M15 (blue) and B10 (orange) models using the NGC\,2516 rotation period observations as the starting point at 150\,Myr. \emph{Top}: Median distance between the slow rotator sequence of NGC\,3532 and the evolved sequence of NGC\,2516 against age. Coloured points show the measured distance and the solid black lines are the fitted polynomials. Their minima (219\,Myr and 265\,Myr) are marked by the solid vertical lines. The age uncertainties (18\,Myr and 26\,Myr) are indicated by the vertical dotted lines. The dashed horizontal line for each model marks the sum of minimum of the polynomial and the standard deviation of the residuals to illustrate the calculation of the age uncertainty.
		\emph{Bottom:} Residuals of the fits with their standard deviations are marked using dashed lines.}
	\label{fig:gyroage}
\end{figure}

\begin{figure*}
	\includegraphics[width=\textwidth]{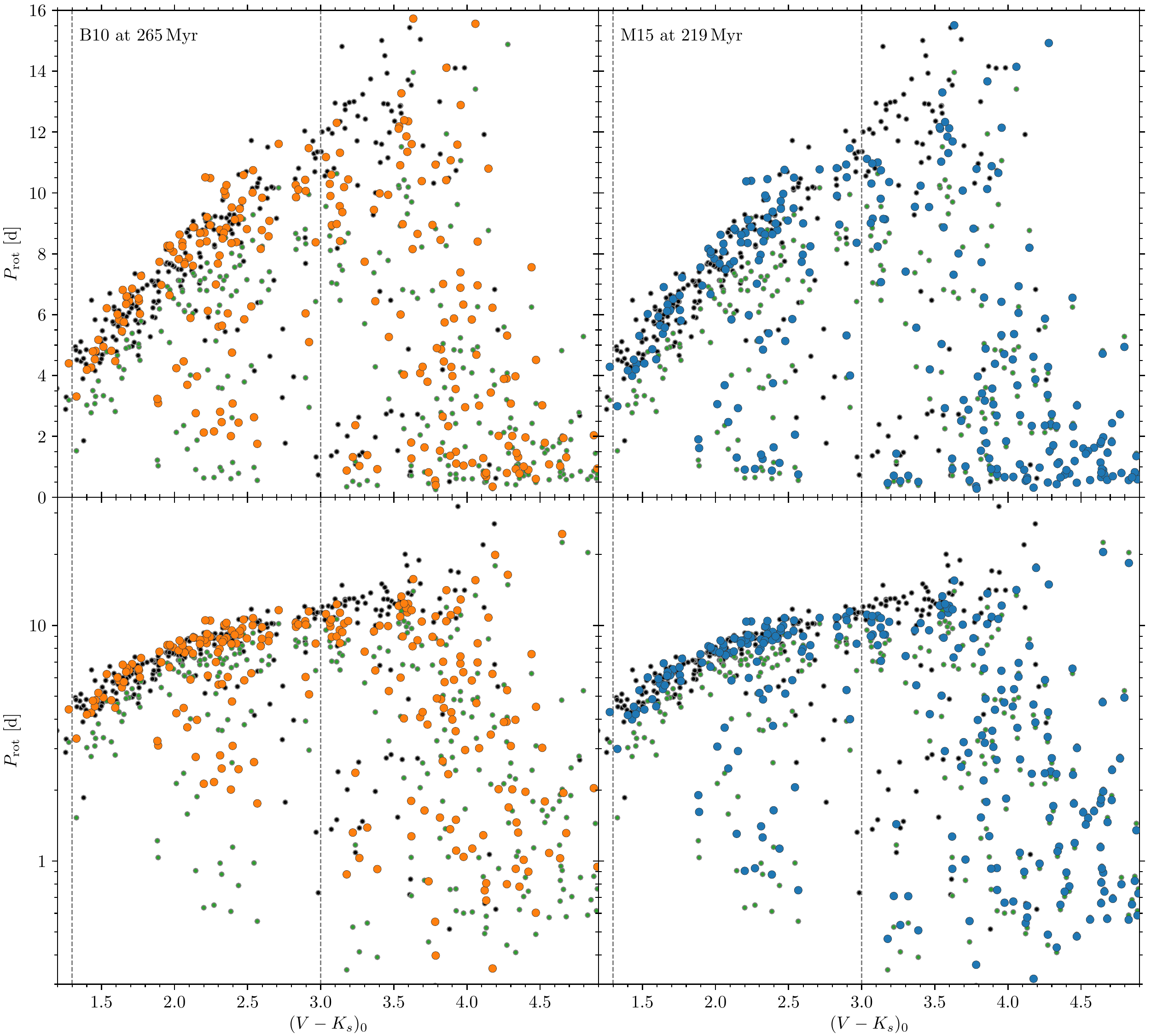}
	\caption{Comparison between the evolved rotation periods (orange and blue) with the observed NGC\,3532 distribution (black). The starting distribution is that of NGC\,2516 (green) and the models aim to match the observed NGC\,3532 rotation periods.
	\emph{Left:} Evolution to 265\,Myr with the B10 model. We see that, although the fast rotators in the NGC\,2516 distribution spin down significantly, the spin-down in the B10 model is still not aggressive enough to transform them to the fast rotator sequence of NGC\,3532 or sufficiently close to it.
	\emph{Right:} Evolution to 219\,Myr with the M15 model. The transition from fast to slow rotation seems to be even  more strongly underestimated, with the fast rotators of the NGC\,2516 distribution not having evolved significantly upwards at all. We note that the lowest-mass stars of the NGC\,2516 sample, which are still on the pre-main sequence, have spun up due to contraction.
	The vertical dashed lines in all panels indicate the colour range used to calculate the best-fitting slow rotator model. In this range both models represent the slow rotators of the target distribution equally well. The \emph{lower} panels (logarithmic in period) feature the full period range, while the \emph{upper} panels are restricted $P_\mathrm{rot}<16$\,d.}
	\label{fig:evolveCPD}
\end{figure*}

A visual comparison of the NGC\,2516 distribution evolved to an age of 300\,Myr with the observed NGC\,3532 distribution confirms our expectations that both rotational evolution models are over-aggressive.
We therefore ask whether these models can reproduce the observed NGC\,3532 distribution when the age constraint is relaxed, and if so, what the relevant age would be for each of the models.
As noted before, an age-offset between the isochronal and the rotational age is not a problem per se, and could potentially be used to adjust the rate of spin-down if the evolved rotation period distribution agrees with the observations. 
This would tell us that the mass dependence of the spin-down formulation is still substantially correct.
Hence, we seek the best-fitting age to determine this difference. 

For each age step, we calculate the distance between the slow rotator sequences of the evolved distribution and the NGC\,3532 observations. 
As defined and initially described in F20, this distance is measured for each period in the evolved data set to the closest star in the NGC\,3532 observations in a normalised CPD\footnote{The $(V-K_s)_0$ colour index is normalised and restricted to the range of the slow rotator sequence of NGC\,2516 ($1.3<(V-K_s)_0<2.0$). 
In contrast, the rotation periods are normalised to the range of NGC\,3532 within this colour range because this cluster exhibits the longer periods.}.

The median value of all the individual distances is then calculated for each age step.
The distribution of these median distances against the age is shown in Fig.~\ref{fig:gyroage} for both models. 
The distributions have very wide minima which are not clearly defined due to the scatter around the general trend. 
Consequently, the best age might not be found at the absolute minimum of the distribution. 
Therefore, we fitted the data with a polynomial of fourth degree and took its minimum as our gyro-age\footnote{The apparent oscillation around the polynomial in Fig.~\ref{fig:gyroage} might be caused by the split sequence (Sect.~\ref{sec:bifurcation}).}.

To estimate the age uncertainties, we calculated the standard deviations of the residuals (Fig.~\ref{fig:gyroage}, lower panel) and added them to the respective minimum values of the polynomial fits. 
The ages at which the polynomial takes these values (dotted lines in Fig.~\ref{fig:gyroage}) are the lower and upper age bounds. 
The differences with respect to the best-fitting ages are the corresponding uncertainties.

For the B10 model, we find NGC\,3532 to be $265\pm26$\,Myr old, somewhat shy of the canonical isochrone age for the cluster,
assuming 150\,Myr for the age of NGC\,2516. 
Hence, NGC\,3532 is 115\,Myr older
than NGC\,2516. 
For the M15 model, we find a younger rotational age of $219\pm18$\,Myr. 
This is significantly short of the canonical isochrone age for NGC\,3532, and only
about the isochronal age of M\,34, which we showed in Sect.~\ref{sec:OCcompare} to be rotationally significantly younger. 
Hence, we can directly conclude that the spin-down for the slow rotators in both models, especially M15, is too strong, and the corresponding estimated rotational ages too young. 
We show the CPDs with both models in Fig.~\ref{fig:evolveCPD} for the corresponding best-fitting ages of $265$\,Myr (B10) and $219$\,Myr (M15). 

Ignoring the absolute age differences and concentrating on the mass dependence of the spin-down,
both models show good agreement between the evolved periods and the NGC\,3532 slow rotator sequence. 
The linear shape of the slow rotator sequence also seems to be adequately reproduced. 
However, neither the reddest stars on the slow rotator sequence in NGC\,3532 ($(V-K_s)_0 = 3-4$) nor the extended slow rotator sequence are reproduced by the models.

In summary, we find that the B10 and M15 models both overestimate the braking of warmer stars (the latter much more so),
while simultaneously underestimating the braking of the cooler stars.

\subsubsection{(Evolved) fast rotators: Properties and problems}
\label{sec:fastevolution}

The strongest mismatch between the evolved periods and NGC\,3532 is found among the stars in transition from fast to slow rotation. 
The spin-down models underestimate the braking drastically (Fig.~\ref{fig:evolveCPD}), 
in contrast to the slow rotators, where the spin-down in overestimated. 
In order to match the observed 
transitional rotator
periods, we would have to evolve NGC\,2516 up to a model age of ${\sim}400$\,Myr with B10, and even more with M15. 

This increase of age is of course not allowable for the other rotators of the cluster. Two consequences would follow.
Firstly, stars on the slow rotator sequence would spin down to longer rotation periods than observed. 
Secondly, lower-mass stars, which are still on the fast rotator sequence in NGC\,3532, would start to evolve off this sequence.

In particular, the latter effect highlights the mismatch between the models and observations. 
It is rooted in the fact that only stars from a small colour (or mass) range are observed in transition. 
From the CPD in Fig.~\ref{fig:CPD}, we find stars of $2.1 < (V-K_s)_0 < 3.0$ (corresponding to $0.8\gtrsim M_\star/M_\sun \gtrsim 0.7$, \citealt{2013ApJS..208....9P}) with 
$1\,\mathrm{d} < P_\mathrm{rot} < 7\,\mathrm{d}$ 
in a well correlated mass-ranked sequence, showing the sudden and rapid, yet collective transition. 
However, the current angular momentum models propose a more gradual evolution, as seen from the much wider colour, and yet a narrower period range of evolved fast rotators in the forward model (Fig.~\ref{fig:evolveCPD}).

To evaluate the validity of the angular momentum models for the remaining fast rotators, we concentrate on the lower panels of Fig.~\ref{fig:evolveCPD}, where the logarithmic scale emphasises the fast rotators. 
Stars still on the fast rotator sequence ($(V-K_s)_0>3.0$) have been spun down with the B10 model to very similar periods as observed in NGC\,3532. 
Hence, their evolution is accurately described, and we expect the warmer fast rotators to be described correspondingly well at younger ages. 
In contrast, M15 barely spins down the fast rotators and the stars in transition. This effect can also be seen in Fig.~\ref{fig:models} where the B10 fast rotator isochrone lies above the \cite{2019A&A...631A..77A} isochrone (which is an updated version of M15).

In conclusion, we have identified two main problems: 
1) The transition from fast to slow rotation is occurring on timescales shorter   than currently anticipated; stronger braking is needed than in the models, and 
2) The mass-dependence of the transition is much steeper than in the models. 
However, given the stronger spin-down through the rotational gap, this issue might resolve itself because the warm end of the transition sequence reaches the slow rotator sequence earlier, and only a small fraction of stars (in a narrow mass range) is in transition at a given time.

\subsection{Empirical insights into the transition}

\subsubsection{Evolved fast rotators: Empirical constraints on the transition timescale}
\label{sec:timescales}

A short transition timescale is supported not only by the comparison with models but also by the observations from the age-ranked succession of open clusters. 
In Fig.~\ref{fig:cOC}, we compare NGC\,3532 to the ${\sim}80$\,Myr younger M\,34. In this open cluster, stars with $(B-V)_0 \approx 0.8$ (${\sim}0.9M_\sun$) are among the bluest fast rotators, while in NGC\,3532 they have just arrived on the slow rotator sequence. 
Hence, the comparison between the two open clusters gives directly a transition time from fast to slow rotation of $\tau_\mathrm{transition}\approx80$\,Myr (for a $0.9M_\sun$, K2 star).

From investigations of stars in M\,34 \cite{2011ApJ...733..115M} found a transition timescale of 100\,Myr for early K dwarfs. 
This value is in good agreement with our observations. 
A similar transition timescale (of 80\,Myr) was found by \cite{2014ApJ...789..101B} from fits of his model to open cluster data, in particularly M\,34.

Using this transition timescale, one finds a mean spin-down rate during the evolution from fast to slow rotation of
$\Delta P/\Delta t \approx 5\,\mathrm{d}/80\,\mathrm{Myr} = 0.063$\,d\,Myr$^{-1}$. Hence, the maximum braking rate is certainly larger. 
In comparison, the B10 model gives for $\mathrm{Ro}=0.06$ 
-- the position of the rotational gap and the maximum spin-down -- 
$\Delta P/\Delta t=0.019$\,d\,Myr$^{-1}$. 
(For M15 the spin-down rate is about the same as for B10.) 
In conclusion, the spin-down is apparently underestimated by a factor of at least three\footnote{Indeed, \cite{2011ApJ...733..115M} calculated the same values for the spin-down. 
Yet, no theoretical work since has addressed this issue.}.

\subsubsection{Evolved fast rotators: Properties of the transition}

The transition from fast to slow rotation is determined not only by the transition time but also the astrophysical properties that initiate it. 
Currently, two main ideas have been proposed. 
In the deterministic approach, the fast rotators spin down magnetically (with a low braking efficiency) until they attain a certain 
Rossby number 
(e.g. \citealt{2003ApJ...586..464B, 2010ApJ...722..222B, 2015ApJ...799L..23M}). 
At this critical point, the rotation and stellar activity desaturate, meaning either the magnetic field configuration \citep{2003ApJ...586..464B, 2018ApJ...862...90G} or properties which determine the torque \citep{2020ApJ...896..123S} become (strongly) dependent on the rotation rate. 
The probabilistic idea \citep{2014ApJ...789..101B} replaces the gradual approach into the desaturated regime with a probabilistic transition towards stronger spin-down.

Our well-defined transition sequence can help to discriminate between the two possibilities. 
A gradual and common approach of all stars towards the threshold rotation rate would result in a well-defined lower limit of the transition sequence. 
In contrast, a probabilistic regime change would cause a wide spread in periods at a constant colour. 
The period spread of the transition sequence (at a fixed colour) is certainly larger than observed on the slow rotator sequence. 
However, it represents the initial period spread of the fast rotator sequence which is, in relative numbers, much larger than in the slow rotator sequence. Additionally, the sequence is sheared in period direction because of the different braking efficiencies over the whole period range. 
The lower boundary is relatively smooth, favouring a deterministic transition. In conclusion, the tight colour correlation among the stars in transition indicates a deterministic, rather than a probabilistic, regime change. 
Still, our observations do not answer the question of why the transition is deterministic.

\subsubsection{Extended slow rotator sequence: Same spin-down rate as warmer stars}
\label{sec:spindown}

\begin{figure}
	\includegraphics[width=\columnwidth]{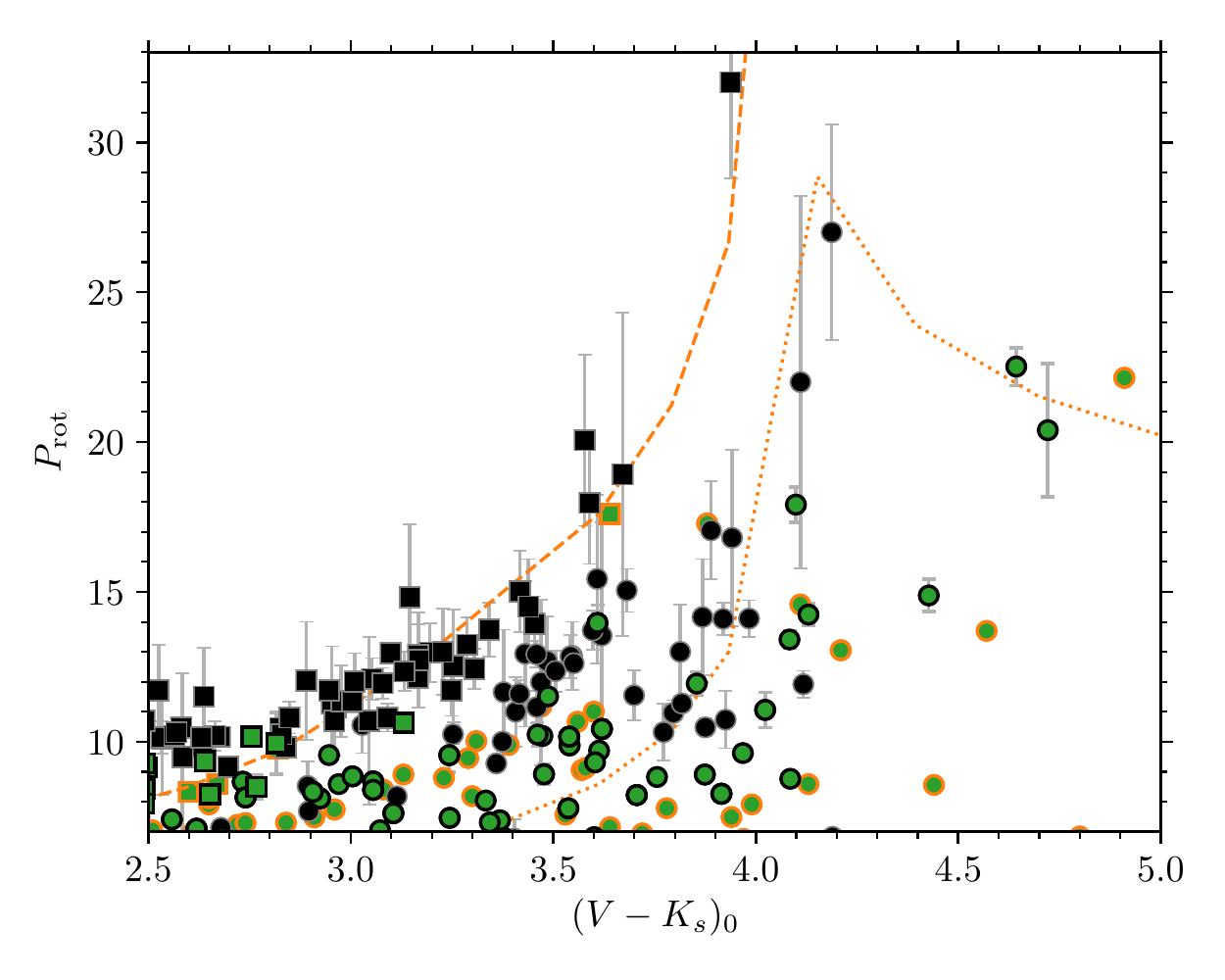}
	\caption{Positions of the extended slow rotator sequences of NGC\,3532 (black) and the coeval Pleiades and NGC\,2516 (green). Both sequences consist of true slow rotators (squares) and gap stars (circles) meaning that each sequence consists of two populations: true slow rotators and stars still in transition. The orange lines are lines of equal Rossby number and describe the position and rotational state of the two populations in the CPD. The dotted line shows $\mathrm{Ro}=0.06$ and describes the position of the rotational gap in \cite{2010ApJ...722..222B}. The orange dashed line is at $\mathrm{Ro}=0.12$, approximately the Rossby number representing the slow rotator sequence in NGC\,3532.	
}
	\label{fig:eSRS}
\end{figure}

The final feature we investigate is the extended slow rotator sequence 
which is the continuation of the slow rotator sequence into the M dwarf regime, with rotation periods up to 32\,d in NGC\,3532. In Fig.~\ref{fig:eSRS}, we show the portion of the CPD with this sequence. 
It includes all three clusters with known extended slow rotator sequences: NGC\,2516 (F20), the Pleiades \citep{2016AJ....152..113R}, and NGC\,3532.
Our identification of the extended slow rotator sequence in NGC\,3532 allows us 
to make the first assessment of the evolution of such stars between {$\sim$}150\,Myr and 300\,Myr. 
From the CPD in Fig.\ref{fig:eSRS}, it becomes clear that these stars actually spin down in that time interval. 
This evidently occurs despite the fact that many of them are on the pre-main sequence at the age of NGC\,2516, and hence still contracting. 
Additionally, the extended slow rotator sequence may actually consist of multiple populations.

In order to investigate the spin-down, we show
lines of equal Rossby number in Fig.~\ref{fig:eSRS}.
The Rossby number, being a mass-normalised rotation period, provides insight into the rotational state of the stars. 
Hence, stars connected by these lines can be considered equivalent in the rotational sense.
Indeed, they tell us that the extended slow rotator sequence consists of true slow rotators with $\textrm{Ro} \approx 0.12$ (similar to their higher mass counterparts) and a group of stars in the rotational gap ($\textrm{Ro} = 0.06$) in transition to the slow rotator sequence.

To estimate the spin down from between the age of the Pleiades and NGC\,3532, we have to assume that we observe the same groups in both open clusters. 
For $(V-K_s)_0 \approx 4.15$, we find a large group of stars in the younger open clusters with $P_\mathrm{rot} \approx 15\,d$. 
In NGC\,3532 two stars with $P_\mathrm{rot}\approx25\,d$ are observed. 
If these two groups of stars correspond to each other at the different ages, than it follows that $\Delta P/\Delta t = 10\,\mathrm{d}/150\,\mathrm{Myr} = 0.067$\,d\,Myr$^{-1}$. 
Unfortunately, the higher mass stars cannot unambiguously be separated. 
Yet, the approximate spin down for the early M dwarfs is the same value as we find for the early K dwarfs (Sect.~\ref{sec:timescales}). 
Hence, we conclude that the spin down rate through the rotational gap is mass independent and only a function of the Rossby number as was proposed in B10.


\section{Conclusions}
\label{sec:conclusion}

Despite the 
ubiquity of fast rotators among cool stars in young open clusters, their evolution and transition to slow rotation has previously not been investigated in detail. 
We do this using the very rich open cluster NGC\,3532, whose membership we determined carefully in our prior study, F19.
In contrast to the plentiful occurrence of ZAMS clusters, NGC\,3532 is uniquely aged for a local cluster, effectively probing the large changes in the rotational evolution of stars between the ZAMS, where rotation has a large spread, and 600\,Myr, where rotation has mostly converged onto a single sequence of period against colour.

In this work, we have described extensive ground-based observations of the 300\,Myr-old open cluster NGC\,3532 with time series photometry to probe this transition.
We determine \npertotal{} rotation periods for members of NGC\,3532 from our photometric time series, spanning stars from late-F to early-M spectral types.
These rotation periods range from 0.5\,d to 32\,d.
While all of these are photometrically determined periods, some of these periods were identified with the help of activity measurements (described in detail in the companion paper, F21act).

The periods delineate a well-populated slow rotator sequence for FGK stars in the colour-period diagram, and a fast rotator sequence for the K dwarfs 
that is significantly evolved beyond that seen in ZAMS-age clusters. 
Among the late K and M dwarfs, we discern both a fast rotator sequence, and an extended slow rotator sequence, the counterpart of the one we (F20) identified in the ZAMS open cluster NGC\,2516.

We observe a split in the slow rotator sequence itself among the warmer stars, one we trace to the single-star vs. binary star sequences in the CMD, allowing us to interpret the observed rotational split as the residual imprint of the spun-down fast rotators. 
In this framework, although the fast rotators spin down rapidly through the rotational gap, they appear not to have fully reached the slow rotator sequence at this age. 

A comparison with
the rotation period distributions of open clusters 
that are nearest in age, M\,34 (220\,Myr) and M\,48 (450\,Myr), shows 
that, as expected, the slow rotator sequence of NGC\,3532 is clearly above that of the former, and below that of the latter, demonstrating that ${\sim} 100$\,Myr age differences between young clusters are visually recognisable in colour-period diagrams.
The empirical intercomparison also highlights the dramatic and mass-dependent 
spin-down of the fast rotators through the rotational gap.

A comparison with rotational isochrones calculated from four prominent models shows detailed issues with reproducing the observed rotational distribution of NGC\,3532.
The mass-dependent dispersion in rotation periods is clearly difficult to achieve, there are issues with the shape of the slow rotator sequence, and the spin-down of the slow rotators is overestimated, while that of the fast rotators is somewhat underestimated.

In an effort to understand the shortcomings of the angular momentum evolution models, we directly evolved the rotation periods of the zero-age main sequence open cluster NGC\,2516 forward in time using two spin-down models, 
avoiding the possible influence of assumptions regarding initial conditions.
These give gyro-ages for NGC\,3532 of $219 \pm 18$\,Myr (M15), and $265 \pm 26$\,Myr (B10), short of the canonical 300\,Myr age for the cluster.
This confirms our prior conclusion that spin-down of slow rotators is overaggressive in rotational models.

The detailed comparison also shows that current models underestimate the spin-down of fast rotators.
The observed spin-down is steeper as a function of stellar mass.
Based on the comparison with the younger cluster M\,34, we estimate a spin-down timescale as short as 80\,Myr and a braking rate $\Delta P/\Delta t \approx 0.06$\,d\,Myr$^{-1}$ for early K\,dwarf fast rotators. 
The shape of the sequence 
suggests that the transition is occurring in a deterministic, rather than a probabilistic, manner.

The sensitivity of our study to long periods enables us to identify in NGC\,3532 the counterpart to the extended slow rotator sequence previously identified by F20 in NGC\,2516.
A similar braking rate is estimated for the slowly rotating M\,dwarfs, taking the age difference between the NGC\,2516/Pleiades and NGC\,3532 as the baseline. 
The rate of spin-down through the rotational gap is apparently not a function of mass, but rather a simple function of the Rossby number, as initially proposed in \cite{2010ApJ...722..222B}. 

In summary, we have performed a comprehensive rotational study of the most populous available open cluster between Pleiades and Hyades age, namely NGC\,3532,
permitting a deeper empirical understanding of the rotational transition between fast rotation at ZAMS age and slow rotation at $600$\,Myr. 
This allows us to place strong mass-dependent constraints on the corresponding requirements for models of stellar rotational evolution.

\begin{acknowledgements}
We are obliged to the referee for a careful and helpful report.
SAB acknowledges support from the Deutsche Forschungs Gemeinschaft (DFG) through project number STR645/7-1.
Based in part on observations at Cerro Tololo Inter-American Observatory, National Optical Astronomy Observatory (2008A-0476; S.~Barnes, SMARTS consortium through Vanderbilt University), which is operated by the Association of Universities for Research in Astronomy (AURA) under a cooperative agreement with the National Science Foundation.
This research has made use of NASA's Astrophysics Data System Bibliographic Services.
This research has made use of the SIMBAD database and the VizieR catalogue access tool, operated at CDS, Strasbourg, France.
This research has made use of NASA's Astrophysics Data System Bibliographic Services.
The Digitized Sky Survey was produced at the Space Telescope Science Institute under U.S. Government grant NAG W-2166. The images of the Digitized Sky Survey are based on photographic data obtained using the Oschin Schmidt Telescope on Palomar Mountain and the UK Schmidt Telescope.
This publication makes use of data products from the Two Micron All Sky Survey, which is a joint project of the University of Massachusetts and the Infrared Processing and Analysis Center/California Institute of Technology, funded by the National Aeronautics and Space Administration and the National Science Foundation.
This work has made use of data from the European Space Agency (ESA) mission
\emph{Gaia} (\url{https://www.cosmos.esa.int/gaia}), processed by the \emph{Gaia}
Data Processing and Analysis Consortium (DPAC,
\url{https://www.cosmos.esa.int/web/gaia/dpac/consortium}). Funding for the DPAC
has been provided by national institutions, in particular the institutions
participating in the \emph{Gaia} Multilateral Agreement.
\newline
\textbf{Software:}
\textsc{DaoPhot~II} was kindly provided by Peter. B. Stetson.
This research made use of \textsc{Astropy}, a community-developed core Python package for Astronomy \citep{2013A&A...558A..33A}.
This work made use of \textsc{Topcat} \citep{2005ASPC..347...29T} and \textsc{Astrometry.net} \citep{astrometry}.
This research made use of the following \textsc{Python} packages:
\textsc{Pandas} \citep{pandas};
\textsc{NumPy} \citep{numpy};
\textsc{MatPlotLib} \citep{Hunter:2007};
\textsc{IPython} \citep{ipython};
\textsc{SciPy} \citep{scipy};
\textsc{Scikit-learn} \citep{scikit-learn}
\end{acknowledgements}

\bibliographystyle{aa} 

\bibliography{NGC3532-full} 

\begin{thebibliography}{95}
\expandafter\ifx\csname natexlab\endcsname\relax\def\natexlab#1{#1}\fi

\bibitem[{{Amard} {et~al.}(2019){Amard}, {Palacios}, {Charbonnel}, {Gallet},
  {Georgy}, {Lagarde}, \& {Siess}}]{2019A&A...631A..77A}
{Amard}, L., {Palacios}, A., {Charbonnel}, C., {et~al.} 2019, \aap, 631, A77

\bibitem[{{Artymowicz} \& {Lubow}(1994)}]{1994ApJ...421..651A}
{Artymowicz}, P. \& {Lubow}, S.~H. 1994, \apj, 421, 651

\bibitem[{{Astropy Collaboration} {et~al.}(2013){Astropy Collaboration},
  {Robitaille}, {Tollerud}, {Greenfield}, {Droettboom}, {Bray}, {Aldcroft},
  {Davis}, {Ginsburg}, {Price-Whelan}, {Kerzendorf}, {Conley}, {Crighton},
  {Barbary}, {Muna}, {Ferguson}, {Grollier}, {Parikh}, {Nair}, {Unther},
  {Deil}, {Woillez}, {Conseil}, {Kramer}, {Turner}, {Singer}, {Fox}, {Weaver},
  {Zabalza}, {Edwards}, {Azalee Bostroem}, {Burke}, {Casey}, {Crawford},
  {Dencheva}, {Ely}, {Jenness}, {Labrie}, {Lim}, {Pierfederici}, {Pontzen},
  {Ptak}, {Refsdal}, {Servillat}, \& {Streicher}}]{2013A&A...558A..33A}
{Astropy Collaboration}, {Robitaille}, T.~P., {Tollerud}, E.~J., {et~al.} 2013,
  \aap, 558, A33

\bibitem[{{Baraffe} {et~al.}(1998){Baraffe}, {Chabrier}, {Allard}, \&
  {Hauschildt}}]{1998A&A...337..403B}
{Baraffe}, I., {Chabrier}, G., {Allard}, F., \& {Hauschildt}, P.~H. 1998, \aap,
  337, 403

\bibitem[{{Baraffe} {et~al.}(2015){Baraffe}, {Homeier}, {Allard}, \&
  {Chabrier}}]{2015A&A...577A..42B}
{Baraffe}, I., {Homeier}, D., {Allard}, F., \& {Chabrier}, G. 2015, \aap, 577,
  A42

\bibitem[{{Barnes}(1997)}]{1997PhDT.........7B}
{Barnes}, S.~A. 1997, PhD thesis, Yale University

\bibitem[{{Barnes}(2003)}]{2003ApJ...586..464B}
{Barnes}, S.~A. 2003, \apj, 586, 464

\bibitem[{{Barnes}(2010)}]{2010ApJ...722..222B}
{Barnes}, S.~A. 2010, \apj, 722, 222

\bibitem[{{Barnes} \& {Kim}(2010)}]{2010ApJ...721..675B}
{Barnes}, S.~A. \& {Kim}, Y.-C. 2010, \apj, 721, 675

\bibitem[{{Barnes} {et~al.}(2016){Barnes}, {Weingrill}, {Fritzewski},
  {Strassmeier}, \& {Platais}}]{2016ApJ...823...16B}
{Barnes}, S.~A., {Weingrill}, J., {Fritzewski}, D., {Strassmeier}, K.~G., \&
  {Platais}, I. 2016, \apj, 823, 16

\bibitem[{{Barnes} {et~al.}(2015){Barnes}, {Weingrill}, {Granzer}, {Spada}, \&
  {Strassmeier}}]{2015A&A...583A..73B}
{Barnes}, S.~A., {Weingrill}, J., {Granzer}, T., {Spada}, F., \& {Strassmeier},
  K.~G. 2015, \aap, 583, A73

\bibitem[{{Belokurov} {et~al.}(2020){Belokurov}, {Penoyre}, {Oh}, {Iorio},
  {Hodgkin}, {Evans}, {Everall}, {Koposov}, {Tout}, {Izzard}, {Clarke}, \&
  {Brown}}]{2020MNRAS.496.1922B}
{Belokurov}, V., {Penoyre}, Z., {Oh}, S., {et~al.} 2020, \mnras, 496, 1922

\bibitem[{{Brown}(2014)}]{2014ApJ...789..101B}
{Brown}, T.~M. 2014, \apj, 789, 101

\bibitem[{{Cargile} {et~al.}(2014){Cargile}, {James}, {Pepper}, {Kuhn},
  {Siverd}, \& {Stassun}}]{2014ApJ...782...29C}
{Cargile}, P.~A., {James}, D.~J., {Pepper}, J., {et~al.} 2014, \apj, 782, 29

\bibitem[{{Casagrande} \& {VandenBerg}(2018)}]{2018MNRAS.479L.102C}
{Casagrande}, L. \& {VandenBerg}, D.~A. 2018, \mnras, 479, L102

\bibitem[{{Chaboyer} {et~al.}(1995){Chaboyer}, {Demarque}, \&
  {Pinsonneault}}]{1995ApJ...441..865C}
{Chaboyer}, B., {Demarque}, P., \& {Pinsonneault}, M.~H. 1995, \apj, 441, 865

\bibitem[{{Cieza} {et~al.}(2009){Cieza}, {Padgett}, {Allen}, {McCabe},
  {Brooke}, {Carey}, {Chapman}, {Fukagawa}, {Huard}, {Noriga-Crespo},
  {Peterson}, \& {Rebull}}]{2009ApJ...696L..84C}
{Cieza}, L.~A., {Padgett}, D.~L., {Allen}, L.~E., {et~al.} 2009, \apjl, 696,
  L84

\bibitem[{{Clem} {et~al.}(2011){Clem}, {Landolt}, {Hoard}, \& {Wachter}}]{clem}
{Clem}, J.~L., {Landolt}, A.~U., {Hoard}, D.~W., \& {Wachter}, S. 2011, \aj,
  141, 115

\bibitem[{{Cohen} {et~al.}(2010){Cohen}, {Drake}, {Kashyap}, {Sokolov}, \&
  {Gombosi}}]{2010ApJ...723L..64C}
{Cohen}, O., {Drake}, J.~J., {Kashyap}, V.~L., {Sokolov}, I.~V., \& {Gombosi},
  T.~I. 2010, \apjl, 723, L64

\bibitem[{{Crane}(2001)}]{2001SoPh..203..381C}
{Crane}, P.~C. 2001, \solphys, 203, 381

\bibitem[{{Cranmer} \& {Saar}(2011)}]{2011ApJ...741...54C}
{Cranmer}, S.~R. \& {Saar}, S.~H. 2011, \apj, 741, 54

\bibitem[{{Curtis} {et~al.}(2019){Curtis}, {Ag{\"u}eros}, {Douglas}, \&
  {Meibom}}]{2019ApJ...879...49C}
{Curtis}, J.~L., {Ag{\"u}eros}, M.~A., {Douglas}, S.~T., \& {Meibom}, S. 2019,
  \apj, 879, 49

\bibitem[{{Delorme} {et~al.}(2011){Delorme}, {Collier Cameron}, {Hebb},
  {Rostron}, {Lister}, {Norton}, {Pollacco}, \& {West}}]{2011MNRAS.413.2218D}
{Delorme}, P., {Collier Cameron}, A., {Hebb}, L., {et~al.} 2011, \mnras, 413,
  2218

\bibitem[{{Douglas} {et~al.}(2016){Douglas}, {Ag{\"u}eros}, {Covey}, {Cargile},
  {Barclay}, {Cody}, {Howell}, \& {Kopytova}}]{2016ApJ...822...47D}
{Douglas}, S.~T., {Ag{\"u}eros}, M.~A., {Covey}, K.~R., {et~al.} 2016, \apj,
  822, 47

\bibitem[{{Douglas} {et~al.}(2019){Douglas}, {Curtis}, {Ag{\"u}eros},
  {Cargile}, {Brewer}, {Meibom}, \& {Jansen}}]{2019ApJ...879..100D}
{Douglas}, S.~T., {Curtis}, J.~L., {Ag{\"u}eros}, M.~A., {et~al.} 2019, \apj,
  879, 100

\bibitem[{{Dworetsky}(1983)}]{1983MNRAS.203..917D}
{Dworetsky}, M.~M. 1983, \mnras, 203, 917

\bibitem[{{Evans} {et~al.}(2018){Evans}, {Riello}, {De Angeli}, {Carrasco},
  {Montegriffo}, {Fabricius}, {Jordi}, {Palaversa}, {Diener}, {Busso},
  {Cacciari}, {van Leeuwen}, {Burgess}, {Davidson}, {Harrison}, {Hodgkin},
  {Pancino}, {Richards}, {Altavilla}, {Balaguer-N{\'u}{\~n}ez}, {Barstow},
  {Bellazzini}, {Brown}, {Castellani}, {Cocozza}, {De Luise}, {Delgado},
  {Ducourant}, {Galleti}, {Gilmore}, {Giuffrida}, {Holl}, {Kewley}, {Koposov},
  {Marinoni}, {Marrese}, {Osborne}, {Piersimoni}, {Portell}, {Pulone},
  {Ragaini}, {Sanna}, {Terrett}, {Walton}, {Wevers}, \&
  {Wyrzykowski}}]{2018A&A...616A...4E}
{Evans}, D.~W., {Riello}, M., {De Angeli}, F., {et~al.} 2018, \aap, 616, A4

\bibitem[{{Fritzewski} {et~al.}(2019){Fritzewski}, {Barnes}, {James}, {Geller},
  {Meibom}, \& {Strassmeier}}]{N3532RV}
{Fritzewski}, D.~J., {Barnes}, S.~A., {James}, D.~J., {et~al.} 2019, \aap, 622,
  A110

\bibitem[{{Fritzewski} {et~al.}({submitted}){Fritzewski}, {Barnes}, {James},
  {Järvinen}, \& {Strassmeier}}]{N3532act}
{Fritzewski}, D.~J., {Barnes}, S.~A., {James}, D.~J., {Järvinen}, S.~P., \&
  {Strassmeier}, K.~G. {submitted}, \aap

\bibitem[{{Fritzewski} {et~al.}(2020){Fritzewski}, {Barnes}, {James}, \&
  {Strassmeier}}]{2020A&A...641A..51F}
{Fritzewski}, D.~J., {Barnes}, S.~A., {James}, D.~J., \& {Strassmeier}, K.~G.
  2020, \aap, 641, A51

\bibitem[{{Gaia Collaboration} {et~al.}(2018){Gaia Collaboration}, {Brown},
  {Vallenari}, {Prusti}, {de Bruijne}, {Babusiaux}, {Bailer-Jones}, {Biermann},
  {Evans}, {Eyer}, {Jansen}, {Jordi}, {Klioner}, {Lammers}, {Lindegren},
  {Luri}, {Mignard}, {Panem}, {Pourbaix}, {Randich}, {Sartoretti}, {Siddiqui},
  {Soubiran}, {van Leeuwen}, {Walton}, {Arenou}, {Bastian}, {Cropper},
  {Drimmel}, {Katz}, {Lattanzi}, {Bakker}, {Cacciari}, {Casta{\~n}eda},
  {Chaoul}, {Cheek}, {De Angeli}, {Fabricius}, {Guerra}, {Holl}, {Masana},
  {Messineo}, {Mowlavi}, {Nienartowicz}, {Panuzzo}, {Portell}, {Riello},
  {Seabroke}, {Tanga}, {Th{\'e}venin}, {Gracia-Abril}, {Comoretto},
  {Garcia-Reinaldos}, {Teyssier}, {Altmann}, {Andrae}, {Audard},
  {Bellas-Velidis}, {Benson}, {Berthier}, {Blomme}, {Burgess}, {Busso},
  {Carry}, {Cellino}, {Clementini}, {Clotet}, {Creevey}, {Davidson}, {De
  Ridder}, {Delchambre}, {Dell'Oro}, {Ducourant}, {Fern{\'a}ndez-
  Hern{\'a}ndez}, {Fouesneau}, {Fr{\'e}mat}, {Galluccio}, {Garc{\'\i}a-Torres},
  {Gonz{\'a}lez-N{\'u}{\~n}ez}, {Gonz{\'a}lez-Vidal}, {Gosset}, {Guy},
  {Halbwachs}, {Hambly}, {Harrison}, {Hern{\'a}ndez}, {Hestroffer}, {Hodgkin},
  {Hutton}, {Jasniewicz}, {Jean-Antoine-Piccolo}, {Jordan}, {Korn},
  {Krone-Martins}, {Lanzafame}, {Lebzelter}, {L{\"o}ffler}, {Manteiga},
  {Marrese}, {Mart{\'\i}n-Fleitas}, {Moitinho}, {Mora}, {Muinonen}, {Osinde},
  {Pancino}, {Pauwels}, {Petit}, {Recio-Blanco}, {Richards}, {Rimoldini},
  {Robin}, {Sarro}, {Siopis}, {Smith}, {Sozzetti}, {S{\"u}veges}, {Torra}, {van
  Reeven}, {Abbas}, {Abreu Aramburu}, {Accart}, {Aerts}, {Altavilla},
  {{\'A}lvarez}, {Alvarez}, {Alves}, {Anderson}, {Andrei}, {Anglada Varela},
  {Antiche}, {Antoja}, {Arcay}, {Astraatmadja}, {Bach}, {Baker},
  {Balaguer-N{\'u}{\~n}ez}, {Balm}, {Barache}, {Barata}, {Barbato}, {Barblan},
  {Barklem}, {Barrado}, {Barros}, {Barstow}, {Bartholom{\'e} Mu{\~n}oz},
  {Bassilana}, {Becciani}, {Bellazzini}, {Berihuete}, {Bertone}, {Bianchi},
  {Bienaym{\'e}}, {Blanco-Cuaresma}, {Boch}, {Boeche}, {Bombrun}, {Borrachero},
  {Bossini}, {Bouquillon}, {Bourda}, {Bragaglia}, {Bramante}, {Breddels},
  {Bressan}, {Brouillet}, {Br{\"u}semeister}, {Brugaletta}, {Bucciarelli},
  {Burlacu}, {Busonero}, {Butkevich}, {Buzzi}, {Caffau}, {Cancelliere},
  {Cannizzaro}, {Cantat-Gaudin}, {Carballo}, {Carlucci}, {Carrasco},
  {Casamiquela}, {Castellani}, {Castro-Ginard}, {Charlot}, {Chemin},
  {Chiavassa}, {Cocozza}, {Costigan}, {Cowell}, {Crifo}, {Crosta}, {Crowley},
  {Cuypers}, {Dafonte}, {Damerdji}, {Dapergolas}, {David}, {David}, {de
  Laverny}, {De Luise}, {De March}, {de Martino}, {de Souza}, {de Torres},
  {Debosscher}, {del Pozo}, {Delbo}, {Delgado}, {Delgado}, {Di Matteo},
  {Diakite}, {Diener}, {Distefano}, {Dolding}, {Drazinos}, {Dur{\'a}n},
  {Edvardsson}, {Enke}, {Eriksson}, {Esquej}, {Eynard Bontemps}, {Fabre},
  {Fabrizio}, {Faigler}, {Falc{\~a}o}, {Farr{\`a}s Casas}, {Federici},
  {Fedorets}, {Fernique}, {Figueras}, {Filippi}, {Findeisen}, {Fonti},
  {Fraile}, {Fraser}, {Fr{\'e}zouls}, {Gai}, {Galleti}, {Garabato},
  {Garc{\'\i}a-Sedano}, {Garofalo}, {Garralda}, {Gavel}, {Gavras}, {Gerssen},
  {Geyer}, {Giacobbe}, {Gilmore}, {Girona}, {Giuffrida}, {Glass}, {Gomes},
  {Granvik}, {Gueguen}, {Guerrier}, {Guiraud}, {Guti{\'e}rrez-S{\'a}nchez},
  {Haigron}, {Hatzidimitriou}, {Hauser}, {Haywood}, {Heiter}, {Helmi}, {Heu},
  {Hilger}, {Hobbs}, {Hofmann}, {Holland}, {Huckle}, {Hypki}, {Icardi},
  {Jan{\ss}en}, {Jevardat de Fombelle}, {Jonker}, {Juh{\'a}sz}, {Julbe},
  {Karampelas}, {Kewley}, {Klar}, {Kochoska}, {Kohley}, {Kolenberg},
  {Kontizas}, {Kontizas}, {Koposov}, {Kordopatis}, {Kostrzewa-Rutkowska},
  {Koubsky}, {Lambert}, {Lanza}, {Lasne}, {Lavigne}, {Le Fustec}, {Le
  Poncin-Lafitte}, {Lebreton}, {Leccia}, {Leclerc}, {Lecoeur-Taibi},
  {Lenhardt}, {Leroux}, {Liao}, {Licata}, {Lindstr{\o}m}, {Lister}, {Livanou},
  {Lobel}, {L{\'o}pez}, {Managau}, {Mann}, {Mantelet}, {Marchal}, {Marchant},
  {Marconi}, {Marinoni}, {Marschalk{\'o}}, {Marshall}, {Martino}, {Marton},
  {Mary}, {Massari}, {Matijevi{\v{c}}}, {Mazeh}, {McMillan}, {Messina},
  {Michalik}, {Millar}, {Molina}, {Molinaro}, {Moln{\'a}r}, {Montegriffo},
  {Mor}, {Morbidelli}, {Morel}, {Morris}, {Mulone}, {Muraveva}, {Musella},
  {Nelemans}, {Nicastro}, {Noval}, {O'Mullane}, {Ord{\'e}novic},
  {Ord{\'o}{\~n}ez-Blanco}, {Osborne}, {Pagani}, {Pagano}, {Pailler},
  {Palacin}, {Palaversa}, {Panahi}, {Pawlak}, {Piersimoni}, {Pineau}, {Plachy},
  {Plum}, {Poggio}, {Poujoulet}, {Pr{\v{s}}a}, {Pulone}, {Racero}, {Ragaini},
  {Rambaux}, {Ramos-Lerate}, {Regibo}, {Reyl{\'e}}, {Riclet}, {Ripepi}, {Riva},
  {Rivard}, {Rixon}, {Roegiers}, {Roelens}, {Romero-G{\'o}mez}, {Rowell},
  {Royer}, {Ruiz-Dern}, {Sadowski}, {Sagrist{\`a} Sell{\'e}s}, {Sahlmann},
  {Salgado}, {Salguero}, {Sanna}, {Santana- Ros}, {Sarasso}, {Savietto},
  {Schultheis}, {Sciacca}, {Segol}, {Segovia}, {S{\'e}gransan}, {Shih},
  {Siltala}, {Silva}, {Smart}, {Smith}, {Solano}, {Solitro}, {Sordo}, {Soria
  Nieto}, {Souchay}, {Spagna}, {Spoto}, {Stampa}, {Steele},
  {Steidelm{\"u}ller}, {Stephenson}, {Stoev}, {Suess}, {Surdej}, {Szabados},
  {Szegedi-Elek}, {Tapiador}, {Taris}, {Tauran}, {Taylor}, {Teixeira},
  {Terrett}, {Teyssandier}, {Thuillot}, {Titarenko}, {Torra Clotet}, {Turon},
  {Ulla}, {Utrilla}, {Uzzi}, {Vaillant}, {Valentini}, {Valette}, {van Elteren},
  {Van Hemelryck}, {van Leeuwen}, {Vaschetto}, {Vecchiato}, {Veljanoski},
  {Viala}, {Vicente}, {Vogt}, {von Essen}, {Voss}, {Votruba}, {Voutsinas},
  {Walmsley}, {Weiler}, {Wertz}, {Wevers}, {Wyrzykowski}, {Yoldas},
  {{\v{Z}}erjal}, {Ziaeepour}, {Zorec}, {Zschocke}, {Zucker}, {Zurbach}, \&
  {Zwitter}}]{2018A&A...616A...1G}
{Gaia Collaboration}, {Brown}, A.~G.~A., {Vallenari}, A., {et~al.} 2018, \aap,
  616, A1

\bibitem[{{Garraffo} {et~al.}(2018){Garraffo}, {Drake}, {Dotter}, {Choi},
  {Burke}, {Moschou}, {Alvarado-G{\'o}mez}, {Kashyap}, \&
  {Cohen}}]{2018ApJ...862...90G}
{Garraffo}, C., {Drake}, J.~J., {Dotter}, A., {et~al.} 2018, \apj, 862, 90

\bibitem[{{Gillen} {et~al.}(2020){Gillen}, {Briegal}, {Hodgkin},
  {Foreman-Mackey}, {Van Leeuwen}, {Jackman}, {McCormac}, {West}, {Queloz},
  {Bayliss}, {Goad}, {Watson}, {Wheatley}, {Belardi}, {Burleigh}, {Casewell},
  {Jenkins}, {Raynard}, {Smith}, {Tilbrook}, \& {Vines}}]{2020MNRAS.492.1008G}
{Gillen}, E., {Briegal}, J.~T., {Hodgkin}, S.~T., {et~al.} 2020, \mnras, 492,
  1008

\bibitem[{{Gregory}(1999)}]{1999ApJ...520..361G}
{Gregory}, P.~C. 1999, \apj, 520, 361

\bibitem[{{Gregory} \& {Loredo}(1992)}]{1992ApJ...398..146G}
{Gregory}, P.~C. \& {Loredo}, T.~J. 1992, \apj, 398, 146

\bibitem[{{Gruner} \& {Barnes}(2020)}]{2020A&A...644A..16G}
{Gruner}, D. \& {Barnes}, S.~A. 2020, \aap, 644, A16

\bibitem[{{Hartman} {et~al.}(2010){Hartman}, {Bakos}, {Kov{\'a}cs}, \&
  {Noyes}}]{2010MNRAS.408..475H}
{Hartman}, J.~D., {Bakos}, G.~{\'A}., {Kov{\'a}cs}, G., \& {Noyes}, R.~W. 2010,
  \mnras, 408, 475

\bibitem[{{Hartman} {et~al.}(2008){Hartman}, {Gaudi}, {Holman}, {McLeod},
  {Stanek}, {Barranco}, {Pinsonneault}, \& {Kalirai}}]{2008ApJ...675.1254H}
{Hartman}, J.~D., {Gaudi}, B.~S., {Holman}, M.~J., {et~al.} 2008, \apj, 675,
  1254

\bibitem[{{Healy} \& {McCullough}(2020)}]{2020ApJ...903...99H}
{Healy}, B.~F. \& {McCullough}, P.~R. 2020, \apj, 903, 99

\bibitem[{{Herbst} \& {Mundt}(2005)}]{2005ApJ...633..967H}
{Herbst}, W. \& {Mundt}, R. 2005, \apj, 633, 967

\bibitem[{Hunter(2007)}]{Hunter:2007}
Hunter, J.~D. 2007, Computing in Science \& Engineering, 9, 90

\bibitem[{{Irwin} {et~al.}(2009){Irwin}, {Aigrain}, {Bouvier}, {Hebb},
  {Hodgkin}, {Irwin}, \& {Moraux}}]{2009MNRAS.392.1456I}
{Irwin}, J., {Aigrain}, S., {Bouvier}, J., {et~al.} 2009, \mnras, 392, 1456

\bibitem[{{Irwin} {et~al.}(2006){Irwin}, {Aigrain}, {Hodgkin}, {Irwin},
  {Bouvier}, {Clarke}, {Hebb}, \& {Moraux}}]{2006MNRAS.370..954I}
{Irwin}, J., {Aigrain}, S., {Hodgkin}, S., {et~al.} 2006, \mnras, 370, 954

\bibitem[{{Irwin} {et~al.}(2007){Irwin}, {Hodgkin}, {Aigrain}, {Hebb},
  {Bouvier}, {Clarke}, {Moraux}, \& {Bramich}}]{2007MNRAS.377..741I}
{Irwin}, J., {Hodgkin}, S., {Aigrain}, S., {et~al.} 2007, \mnras, 377, 741

\bibitem[{{James} {et~al.}(2010){James}, {Barnes}, {Meibom}, {Lockwood},
  {Levine}, {Deliyannis}, {Platais}, {Steinhauer}, \&
  {Hurley}}]{2010A&A...515A.100J}
{James}, D.~J., {Barnes}, S.~A., {Meibom}, S., {et~al.} 2010, \aap, 515, A100

\bibitem[{{Johnstone} {et~al.}(2015){Johnstone}, {G{\"u}del}, {St{\"o}kl},
  {Lammer}, {Tu}, {Kislyakova}, {L{\"u}ftinger}, {Odert}, {Erkaev}, \&
  {Dorfi}}]{2015ApJ...815L..12J}
{Johnstone}, C.~P., {G{\"u}del}, M., {St{\"o}kl}, A., {et~al.} 2015, \apjl,
  815, L12

\bibitem[{Jones {et~al.}(2001)Jones, Oliphant, Peterson, {et~al.}}]{scipy}
Jones, E., Oliphant, T., Peterson, P., {et~al.} 2001, {SciPy}: Open source
  scientific tools for {Python}

\bibitem[{{Kawaler}(1988)}]{1988ApJ...333..236K}
{Kawaler}, S.~D. 1988, \apj, 333, 236

\bibitem[{{Kovalev} {et~al.}(2019){Kovalev}, {Bergemann}, {Ting}, \&
  {Rix}}]{2019A&A...628A..54K}
{Kovalev}, M., {Bergemann}, M., {Ting}, Y.-S., \& {Rix}, H.-W. 2019, \aap, 628,
  A54

\bibitem[{{Kraft}(1967)}]{1967ApJ...150..551K}
{Kraft}, R.~P. 1967, \apj, 150, 551

\bibitem[{{Kron}(1947)}]{1947PASP...59..261K}
{Kron}, G.~E. 1947, \pasp, 59, 261

\bibitem[{{Lang} {et~al.}(2010){Lang}, {Hogg}, {Mierle}, {Blanton}, \&
  {Roweis}}]{astrometry}
{Lang}, D., {Hogg}, D.~W., {Mierle}, K., {Blanton}, M., \& {Roweis}, S. 2010,
  \aj, 137, 1782

\bibitem[{{Li} {et~al.}(2020){Li}, {Shao}, {Li}, {Yu}, {Zhong}, \&
  {Chen}}]{2020ApJ...901...49L}
{Li}, L., {Shao}, Z., {Li}, Z.-Z., {et~al.} 2020, \apj, 901, 49

\bibitem[{{Lindegren} {et~al.}(2021){Lindegren}, {Klioner}, {Hern{\'a}ndez},
  {Bombrun}, {Ramos-Lerate}, {Steidelm{\"u}ller}, {Bastian}, {Biermann}, {de
  Torres}, {Gerlach}, {Geyer}, {Hilger}, {Hobbs}, {Lammers}, {McMillan},
  {Stephenson}, {Casta{\~n}eda}, {Davidson}, {Fabricius}, {Gracia-Abril},
  {Portell}, {Rowell}, {Teyssier}, {Torra}, {Bartolom{\'e}}, {Clotet},
  {Garralda}, {Gonz{\'a}lez-Vidal}, {Torra}, {Abbas}, {Altmann}, {Anglada
  Varela}, {Balaguer-N{\'u}{\~n}ez}, {Balog}, {Barache}, {Becciani}, {Bernet},
  {Bertone}, {Bianchi}, {Bouquillon}, {Brown}, {Bucciarelli}, {Busonero},
  {Butkevich}, {Buzzi}, {Cancelliere}, {Carlucci}, {Charlot}, {Cioni},
  {Crosta}, {Crowley}, {del Peloso}, {del Pozo}, {Drimmel}, {Esquej}, {Fienga},
  {Fraile}, {Gai}, {Garcia-Reinaldos}, {Guerra}, {Hambly}, {Hauser},
  {Jan{\ss}en}, {Jordan}, {Kostrzewa-Rutkowska}, {Lattanzi}, {Liao}, {Licata},
  {Lister}, {L{\"o}ffler}, {Marchant}, {Masip}, {Mignard}, {Mints}, {Molina},
  {Mora}, {Morbidelli}, {Murphy}, {Pagani}, {Panuzzo}, {Pe{\~n}alosa Esteller},
  {Poggio}, {Re Fiorentin}, {Riva}, {Sagrist{\`a} Sell{\'e}s}, {Sanchez
  Gimenez}, {Sarasso}, {Sciacca}, {Siddiqui}, {Smart}, {Souami}, {Spagna},
  {Steele}, {Taris}, {Utrilla}, {van Reeven}, \&
  {Vecchiato}}]{2021A&A...649A...2L}
{Lindegren}, L., {Klioner}, S.~A., {Hern{\'a}ndez}, J., {et~al.} 2021, \aap,
  649, A2

\bibitem[{{Lomb}(1976)}]{1976Ap&SS..39..447L}
{Lomb}, N.~R. 1976, \apss, 39, 447

\bibitem[{{Lorenzo-Oliveira} {et~al.}(2020){Lorenzo-Oliveira}, {Mel{\'e}ndez},
  {Ponte}, \& {Galarza}}]{2020MNRAS.495L..61L}
{Lorenzo-Oliveira}, D., {Mel{\'e}ndez}, J., {Ponte}, G., \& {Galarza}, J.~Y.
  2020, \mnras, 495, L61

\bibitem[{{MacGregor} \& {Brenner}(1991)}]{1991ApJ...376..204M}
{MacGregor}, K.~B. \& {Brenner}, M. 1991, \apj, 376, 204

\bibitem[{{Mamajek} \& {Hillenbrand}(2008)}]{2008ApJ...687.1264M}
{Mamajek}, E.~E. \& {Hillenbrand}, L.~A. 2008, \apj, 687, 1264

\bibitem[{{Mathieu}(1994)}]{1994ARA&A..32..465M}
{Mathieu}, R.~D. 1994, \araa, 32, 465

\bibitem[{{Matt} {et~al.}(2015){Matt}, {Brun}, {Baraffe}, {Bouvier}, \&
  {Chabrier}}]{2015ApJ...799L..23M}
{Matt}, S.~P., {Brun}, A.~S., {Baraffe}, I., {Bouvier}, J., \& {Chabrier}, G.
  2015, \apj, 799, L23

\bibitem[{{Matt} {et~al.}(2019){Matt}, {Brun}, {Baraffe}, {Bouvier}, \&
  {Chabrier}}]{2019ApJ...870L..27M}
{Matt}, S.~P., {Brun}, A.~S., {Baraffe}, I., {Bouvier}, J., \& {Chabrier}, G.
  2019, \apjl, 870, L27

\bibitem[{McKinney(2010)}]{pandas}
McKinney, W. 2010, in Proceedings of the 9th Python in Science Conference, ed.
  S.~van~der Walt \& J.~Millman, 51

\bibitem[{{Meibom} {et~al.}(2015){Meibom}, {Barnes}, {Platais}, {Gilliland},
  {Latham}, \& {Mathieu}}]{2015Natur.517..589M}
{Meibom}, S., {Barnes}, S.~A., {Platais}, I., {et~al.} 2015, \nat, 517, 589

\bibitem[{{Meibom} {et~al.}(2007){Meibom}, {Mathieu}, \&
  {Stassun}}]{2007ApJ...665L.155M}
{Meibom}, S., {Mathieu}, R.~D., \& {Stassun}, K.~G. 2007, \apjl, 665, L155

\bibitem[{{Meibom} {et~al.}(2009){Meibom}, {Mathieu}, \&
  {Stassun}}]{2009ApJ...695..679M}
{Meibom}, S., {Mathieu}, R.~D., \& {Stassun}, K.~G. 2009, \apj, 695, 679

\bibitem[{{Meibom} {et~al.}(2011){Meibom}, {Mathieu}, {Stassun}, {Liebesny}, \&
  {Saar}}]{2011ApJ...733..115M}
{Meibom}, S., {Mathieu}, R.~D., {Stassun}, K.~G., {Liebesny}, P., \& {Saar},
  S.~H. 2011, \apj, 733, 115

\bibitem[{{Messina}(2019)}]{2019A&A...627A..97M}
{Messina}, S. 2019, \aap, 627, A97

\bibitem[{{Pecaut} \& {Mamajek}(2013)}]{2013ApJS..208....9P}
{Pecaut}, M.~J. \& {Mamajek}, E.~E. 2013, \apjs, 208, 9

\bibitem[{Pedregosa {et~al.}(2011)Pedregosa, Varoquaux, Gramfort, Michel,
  Thirion, Grisel, Blondel, Prettenhofer, Weiss, Dubourg, Vanderplas, Passos,
  Cournapeau, Brucher, Perrot, \& Duchesnay}]{scikit-learn}
Pedregosa, F., Varoquaux, G., Gramfort, A., {et~al.} 2011, Journal of Machine
  Learning Research, 12, 2825

\bibitem[{Pérez \& Granger(2007)}]{ipython}
Pérez, F. \& Granger, B.~E. 2007, Computing in Science \& Engineering, 9, 21

\bibitem[{{Radick} {et~al.}(1987){Radick}, {Thompson}, {Lockwood}, {Duncan}, \&
  {Baggett}}]{1987ApJ...321..459R}
{Radick}, R.~R., {Thompson}, D.~T., {Lockwood}, G.~W., {Duncan}, D.~K., \&
  {Baggett}, W.~E. 1987, \apj, 321, 459

\bibitem[{{Rebull} {et~al.}(2016){Rebull}, {Stauffer}, {Bouvier}, {Cody},
  {Hillenbrand}, {Soderblom}, {Valenti}, {Barrado}, {Bouy}, {Ciardi},
  {Pinsonneault}, {Stassun}, {Micela}, {Aigrain}, {Vrba}, {Somers},
  {Christiansen}, {Gillen}, \& {Collier Cameron}}]{2016AJ....152..113R}
{Rebull}, L.~M., {Stauffer}, J.~R., {Bouvier}, J., {et~al.} 2016, \aj, 152, 113

\bibitem[{{Rebull} {et~al.}(2017){Rebull}, {Stauffer}, {Hillenbrand}, {Cody},
  {Bouvier}, {Soderblom}, {Pinsonneault}, \& {Hebb}}]{2017ApJ...839...92R}
{Rebull}, L.~M., {Stauffer}, J.~R., {Hillenbrand}, L.~A., {et~al.} 2017, \apj,
  839, 92

\bibitem[{{Roberts} {et~al.}(1987){Roberts}, {Lehar}, \&
  {Dreher}}]{1987AJ.....93..968R}
{Roberts}, D.~H., {Lehar}, J., \& {Dreher}, J.~W. 1987, \aj, 93, 968

\bibitem[{{Scargle}(1982)}]{1982ApJ...263..835S}
{Scargle}, J.~D. 1982, \apj, 263, 835

\bibitem[{{See} {et~al.}(2019{\natexlab{a}}){See}, {Matt}, {Finley}, {Folsom},
  {Boro Saikia}, {Donati}, {Fares}, {H{\'e}brard}, {Jardine}, {Jeffers},
  {Marsden}, {Mengel}, {Morin}, {Petit}, {Vidotto}, {Waite}, \& {the BCool
  Collaboration}}]{2019ApJ...886..120S}
{See}, V., {Matt}, S.~P., {Finley}, A.~J., {et~al.} 2019{\natexlab{a}}, \apj,
  886, 120

\bibitem[{{See} {et~al.}(2019{\natexlab{b}}){See}, {Matt}, {Folsom}, {Boro
  Saikia}, {Donati}, {Fares}, {Finley}, {H{\'e}brard}, {Jardine}, {Jeffers},
  {Lehmann}, {Marsden}, {Mengel}, {Morin}, {Petit}, {Vidotto}, {Waite}, \&
  {BCool Collaboration}}]{2019ApJ...876..118S}
{See}, V., {Matt}, S.~P., {Folsom}, C.~P., {et~al.} 2019{\natexlab{b}}, \apj,
  876, 118

\bibitem[{{Shoda} {et~al.}(2020){Shoda}, {Suzuki}, {Matt}, {Cranmer},
  {Vidotto}, {Strugarek}, {See}, {R{\'e}ville}, {Finley}, \&
  {Brun}}]{2020ApJ...896..123S}
{Shoda}, M., {Suzuki}, T.~K., {Matt}, S.~P., {et~al.} 2020, \apj, 896, 123

\bibitem[{{Skrutskie} {et~al.}(2006){Skrutskie}, {Cutri}, {Stiening},
  {Weinberg}, {Schneider}, {Carpenter}, {Beichman}, {Capps}, {Chester},
  {Elias}, {Huchra}, {Liebert}, {Lonsdale}, {Monet}, {Price}, {Seitzer},
  {Jarrett}, {Kirkpatrick}, {Gizis}, {Howard}, {Evans}, {Fowler}, {Fullmer},
  {Hurt}, {Light}, {Kopan}, {Marsh}, {McCallon}, {Tam}, {Van Dyk}, \&
  {Wheelock}}]{2006AJ....131.1163S}
{Skrutskie}, M.~F., {Cutri}, R.~M., {Stiening}, R., {et~al.} 2006, \aj, 131,
  1163

\bibitem[{{Skumanich}(1972)}]{1972ApJ...171..565S}
{Skumanich}, A. 1972, \apj, 171, 565

\bibitem[{{Soderblom} {et~al.}(1993){Soderblom}, {Stauffer}, {MacGregor}, \&
  {Jones}}]{1993ApJ...409..624S}
{Soderblom}, D.~R., {Stauffer}, J.~R., {MacGregor}, K.~B., \& {Jones}, B.~F.
  1993, \apj, 409, 624

\bibitem[{{Spada} \& {Lanzafame}(2020)}]{2020A&A...636A..76S}
{Spada}, F. \& {Lanzafame}, A.~C. 2020, \aap, 636, A76

\bibitem[{{Stauffer} {et~al.}(2018){Stauffer}, {Rebull}, {Cody}, {Hillenbrand},
  {Pinsonneault}, {Barrado}, {Bouvier}, \& {David}}]{2018AJ....156..275S}
{Stauffer}, J., {Rebull}, L.~M., {Cody}, A.~M., {et~al.} 2018, \aj, 156, 275

\bibitem[{{Stauffer} \& {Hartmann}(1987)}]{1987ApJ...318..337S}
{Stauffer}, J.~R. \& {Hartmann}, L.~W. 1987, \apj, 318, 337

\bibitem[{{Stellingwerf}(1978)}]{1978ApJ...224..953S}
{Stellingwerf}, R.~F. 1978, \apj, 224, 953

\bibitem[{{Stetson}(1987)}]{1987PASP...99..191S}
{Stetson}, P.~B. 1987, \pasp, 99, 191

\bibitem[{{Stetson}(1994)}]{1994PASP..106..250S}
{Stetson}, P.~B. 1994, \pasp, 106, 250

\bibitem[{{Stetson} {et~al.}(2003){Stetson}, {Bruntt}, \&
  {Grundahl}}]{2003PASP..115..413S}
{Stetson}, P.~B., {Bruntt}, H., \& {Grundahl}, F. 2003, \pasp, 115, 413

\bibitem[{{Strassmeier}(2009)}]{2009A&ARv..17..251S}
{Strassmeier}, K.~G. 2009, \aapr, 17, 251

\bibitem[{{Taylor}(2005)}]{2005ASPC..347...29T}
{Taylor}, M.~B. 2005, in Astronomical Society of the Pacific Conference Series,
  Vol. 347, Astronomical Data Analysis Software and Systems XIV, ed.
  P.~{Shopbell}, M.~{Britton}, \& R.~{Ebert}, 29

\bibitem[{van~der Walt {et~al.}(2011)van~der Walt, Colbert, \&
  Varoquaux}]{numpy}
van~der Walt, S., Colbert, S.~C., \& Varoquaux, G. 2011, Computing in Science
  \& Engineering, 13, 22

\bibitem[{{van Leeuwen} \& {Alphenaar}(1982)}]{1982Msngr..28...15V}
{van Leeuwen}, F. \& {Alphenaar}, P. 1982, The Messenger, 28, 15

\bibitem[{{van Leeuwen} {et~al.}(1987){van Leeuwen}, {Alphenaar}, \&
  {Meys}}]{1987A&AS...67..483V}
{van Leeuwen}, F., {Alphenaar}, P., \& {Meys}, J.~J.~M. 1987, \aaps, 67, 483

\bibitem[{{Zahn}(1989)}]{1989A&A...220..112Z}
{Zahn}, J.~P. 1989, \aap, 220, 112

\bibitem[{{Zechmeister} \& {K{\"u}rster}(2009)}]{2009A&A...496..577Z}
{Zechmeister}, M. \& {K{\"u}rster}, M. 2009, \aap, 496, 577

\end{thebibliography}

\begin{appendix}
	\section{Light curves}
	
	\begin{figure*}
		\includegraphics[width=\textwidth]{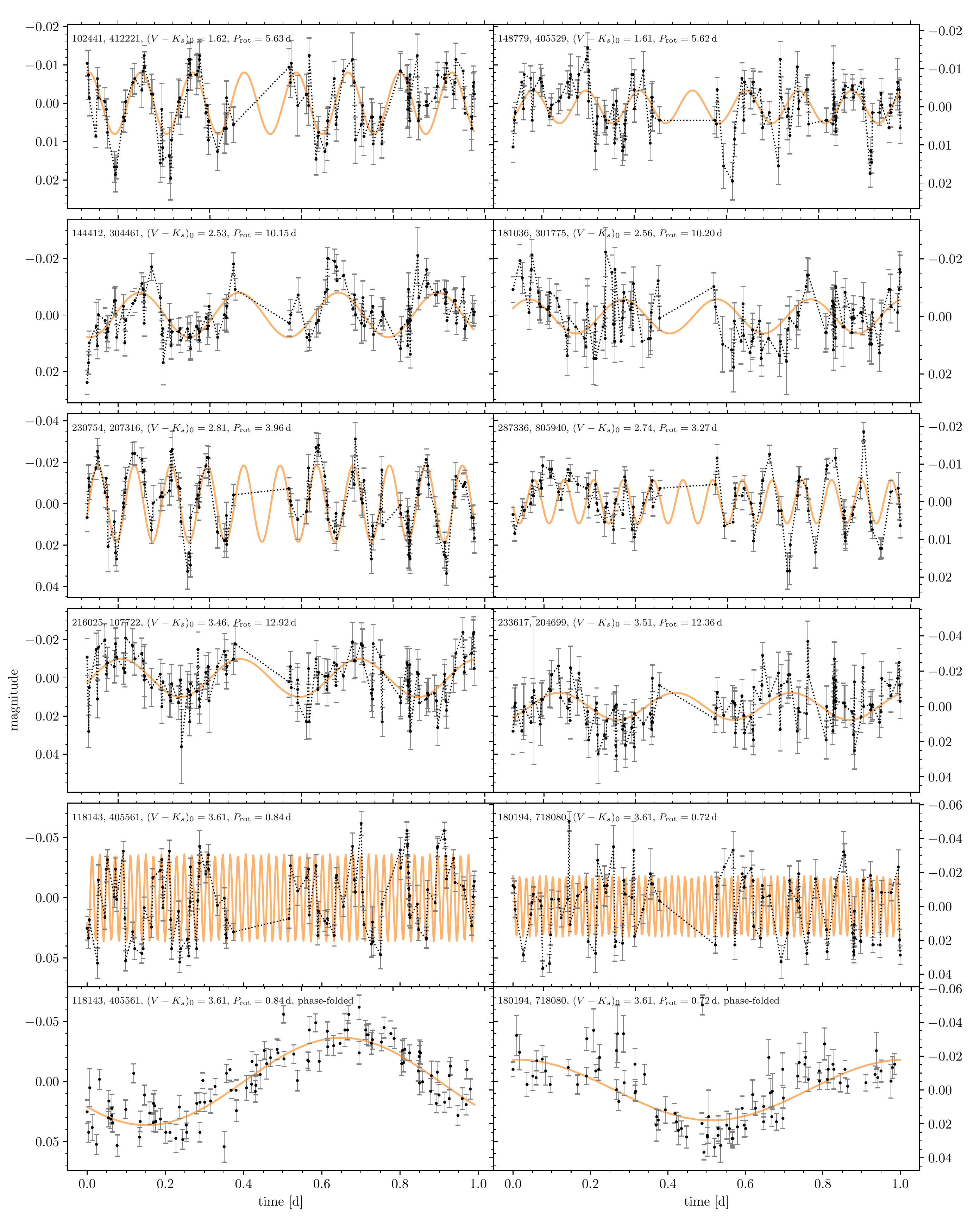}
		\caption{Examples of light curves with measured rotation periods. The stars are also marked in the CPD of Fig.~\ref{fig:CPD}. The \emph{left} column shows light curves classified as first class and the \emph{right} column the noisier algorithmic ones. The bottom row features the phase-folded light curves of the fast rotators in the row immediately above. Each light curve is over-plotted with a sine of the same periodicity as our final rotation period. The ID numbers in each panel are from \cite{clem} and F19.}
		\label{appf:LCs}
	\end{figure*}

\end{appendix}
\end{document}